\documentclass[epj]{svjour}

\usepackage{latexsym}
\usepackage{graphics}
\usepackage{afterpage}
\usepackage{wasysym}

\usepackage{amsfonts}
\usepackage{amssymb}
\usepackage{subfigure}

\usepackage{textcomp}
\usepackage{gensymb}
\usepackage[english]{babel}
\usepackage[T1]{fontenc}
\usepackage{setspace}
\usepackage{color}
\usepackage{siunitx}
\usepackage{isotope}
\usepackage{here}
\usepackage{afterpage}
 
\begin{document}

\title{Solid deuterium surface degradation at ultracold neutron sources}

\author{
A.~Anghel\inst{1},
T.L.~Bailey\inst{2}, 
G.~Bison\inst{1}, 
B.~Blau\inst{1}, 
L.J.~Broussard\inst{3},
S.M.~Clayton\inst{3},
C.~Cude-Woods\inst{3},
M.~Daum\inst{1}, 
A.~Hawari\inst{2},
N.~Hild\inst{1}$^,$\inst{4},
P.~Huffman\inst{2}, 
T.M.~Ito\inst{3},
K.~Kirch\inst{1}$^,$\inst{4},
E.~Korobkina\inst{2}$^,$\thanks{Email:ekorobk@ncsu.edu}, 
B.~Lauss\inst{1}$^,$\thanks{Email:bernhard.lauss@psi.ch},
K.~Leung\inst{2},
E.M.~Lutz\inst{2},  
M.~Makela\inst{3},
G.~Medlin\inst{2}, 
C.L.~Morris\inst{3},
R.W.~Pattie\inst{3},
D.~Ries\inst{5},
A.~Saunders\inst{3},
P.~Schmidt-Wellenburg\inst{1}, 
V.~Talanov\inst{1},
A.R.~Young\inst{2},
B.~Wehring\inst{2},
C.~White\inst{2},
M.~Wohlmuther\inst{1},
G.~Zsigmond\inst{1}$^,$\thanks{Email:geza.zsigmond@psi.ch}
}

\institute{
$^1$Paul Scherrer Institut, CH-5232 Villigen-PSI, Switzerland \\
$^2$North Carolina State University, Raleigh, NC 27695, USA\\
$^3$Los Alamos National Laboratory, Los Alamos, New Mexico 87545, USA\\
$^4$ETH Z\"urich, CH-8093 Z\"urich, Switzerland\\
$^5$Inst. of Nuclear Chemistry,  Johannes Gutenberg University of Mainz, D-55128 Mainz, Germany\\
}


\date{\today}

\abstract{
Solid deuterium (sD$_2$) is used as an
efficient converter to produce ultracold neutrons (UCN).
It is known that the sD$_2$ must be sufficiently cold, of high purity and 
mostly in its ortho-state
in order to guarantee long lifetimes of UCN in the solid from which they 
are extracted into vacuum.
Also the UCN transparency of the bulk sD$_2$ material
must be high because crystal inhomogeneities limit the mean free path for elastic scattering 
and reduce the extraction efficiency.
Observations at the UCN sources at Paul Scherrer Institute and at 
Los Alamos National Laboratory
consistently show a decrease of the UCN yield with time of operation 
after initial preparation 
or later treatment (``conditioning'') of the sD$_2$.
We show that, in addition to the quality of the bulk sD$_2$, 
the quality of its surface is essential.  
Our observations and simulations support
the view that the surface is deteriorating due to 
a build-up of D$_2$ frost-layers under pulsed operation which leads to
strong albedo reflections of UCN and subsequent loss.
We report results of UCN yield measurements, temperature and pressure behavior 
of deuterium during source operation 
and conditioning, and UCN transport simulations. 
This, together with optical observations of sD$_2$ frost formation on initially 
transparent sD$_2$ in offline studies with pulsed heat input 
at the North Carolina State University UCN source results in a consistent description of 
the UCN yield decrease. 
}



%
\authorrunning{Authors et al.}
\titlerunning{D2 frost}

\maketitle





\section{Introduction}

%
Deuterium is
one of the best neutron moderator materials
because of its small mass and small nuclear absorption cross section. 
In the liquid phase, it is the favorite material for cold neutron sources 
at continuous research reactors
and accelerator-based spallation sources. 
As solid deuterium (sD$_2$), it is an exceptional material 
for optimized production of very low energy, 
so-called ``ultracold neutrons (UCN)'',
with energies below about 350\,neV~\cite{Golub1991}. 
These UCN can be trapped for several hundreds of seconds,
limited ultimately by the neutron beta-decay lifetime, allowing
for long observation and manipulation times.
This provides a significant gain for many experiments 
in fundamental, low-energy 
particle physics~\cite{Abele2008,Dubbers2011}.

The quality of a neutron source in terms of the neutron intensity 
delivered and available to experiments
depends on many parameters, including the neutron production process, 
the moderator setup, and the 
specific geometries and materials used.
The potential of sD$_2$ for UCN production in a continuous source 
has been first demonstrated 
experimentally in Gatchina~\cite{Altarev1980b} 
and put on solid theoretical grounds shortly 
thereafter~\cite{Golub1983,Yu1986}. 
In the 1990s, when the saturation of UCN intensities at available sources 
(mainly the Steyerl turbine at the ILL in Grenoble~\cite{Steyerl1986})
became a limiting factor, research into improved UCN production mechanisms
was intensified, concentrating on developing so-called 
super-thermal sources ~\cite{Golub1977}
either using superfluid helium or sD$_2$.

For sD$_2$, pioneering work continued at
the Petersburg Nuclear Physics Institute (PNPI),
Gatchina ~\cite{Serebrov1994,Serebrov1995b,Serebrov2000,Serebrov2001b},
and triggered renewed interest in pulsed sD$_2$ sources,
see~\cite{Pokotilovski1995} for suggestion of a lay-out.
Conceptually, a UCN source would produce neutrons from fission or spallation,
have separate or combined thermal and cold moderation,
and produce UCN inside
a sD$_2$ converter 
that could also serve as the cold moderator.
The UCN produced inside the sD$_2$ must leave the converter into vacuum 
within their lifetime
that is limited by absorption and upscattering losses. 
High UCN densities in vacuum can be obtained by guiding the UCN 
from the converter to a separate, low-loss volume 
that can be closed synchronized to pulsed production.

The first sD$_2$-based UCN source driven by a pulsed proton beam, 
which was realized
in Los Alamos~\cite{Morris2002,Saunders2004} 
has yielded insights
about relevant UCN loss mechanisms inside sD$_2$. 
Thermal and para-D$_2$ 
upscattering~\cite{Morris2002,Liu2000},
were found to limit the lifetime of UCN inside the sD$_2$ 
and affect
efficient UCN extraction from the converter.
In a series of experiments at the Paul Scherrer Insitute (PSI) 
in Villigen and at the Institute Laue-Langevin (ILL), Grenoble,
the UCN production process from cold neutrons (CN) in sD$_2$ 
has been quantitatively studied 
integrated over a CN spectrum~\cite{Atchison2005}, 
CN velocity dependent~\cite{Atchison2007}
and via CN measurements of the sD$_2$ density of states~\cite{Frei2010b}.
The influence of the crystalline properties of the sD$_2$, 
i.e. the effect of UCN elastic scattering on inhomogeneities, 
its variations with quality of the bulk material, 
and consequently the large changes in UCN residence times inside 
and extraction efficiencies out of sD$_2$
into vacuum have been studied in~\cite{Atchison2005b,Atchison2011} 
and further worked out, 
e.g. in~\cite{Pokotilovski2012}.


\begin{figure}[htb]
\begin{center}
\resizebox{0.5\textwidth}{!}{\includegraphics{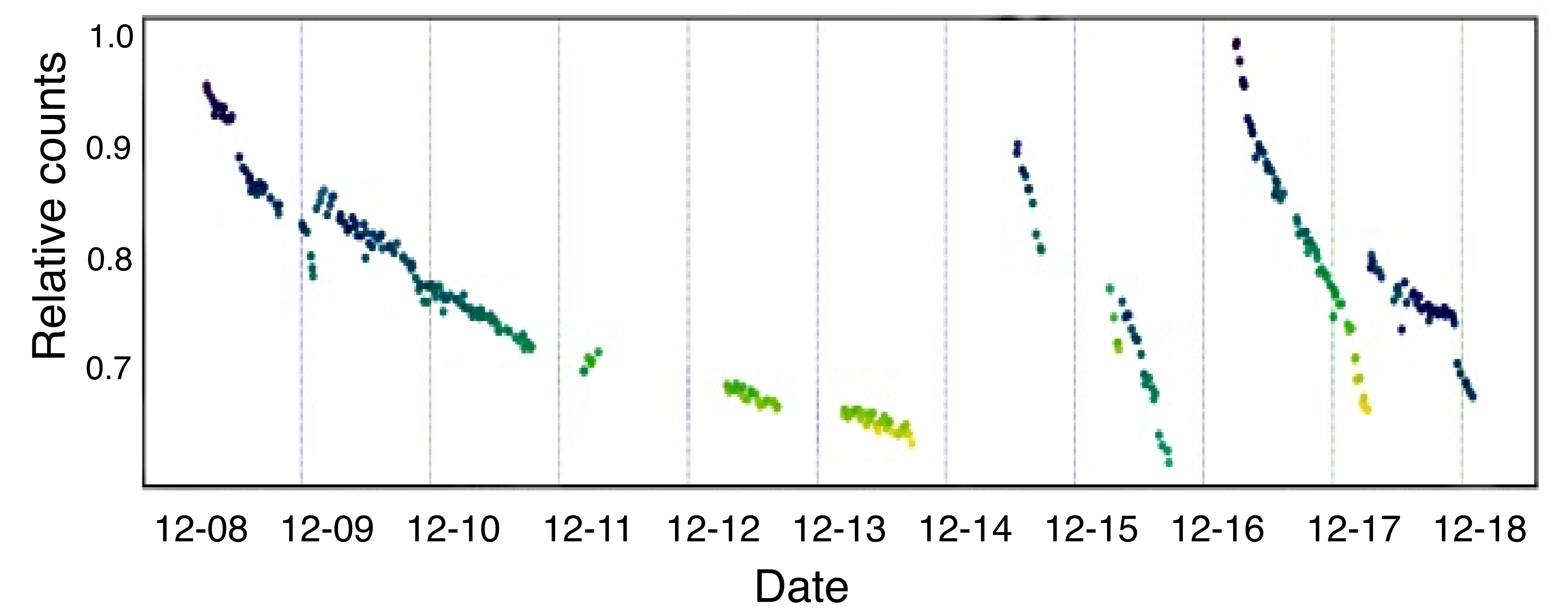}
}
\resizebox{0.5\textwidth}{!}{\includegraphics{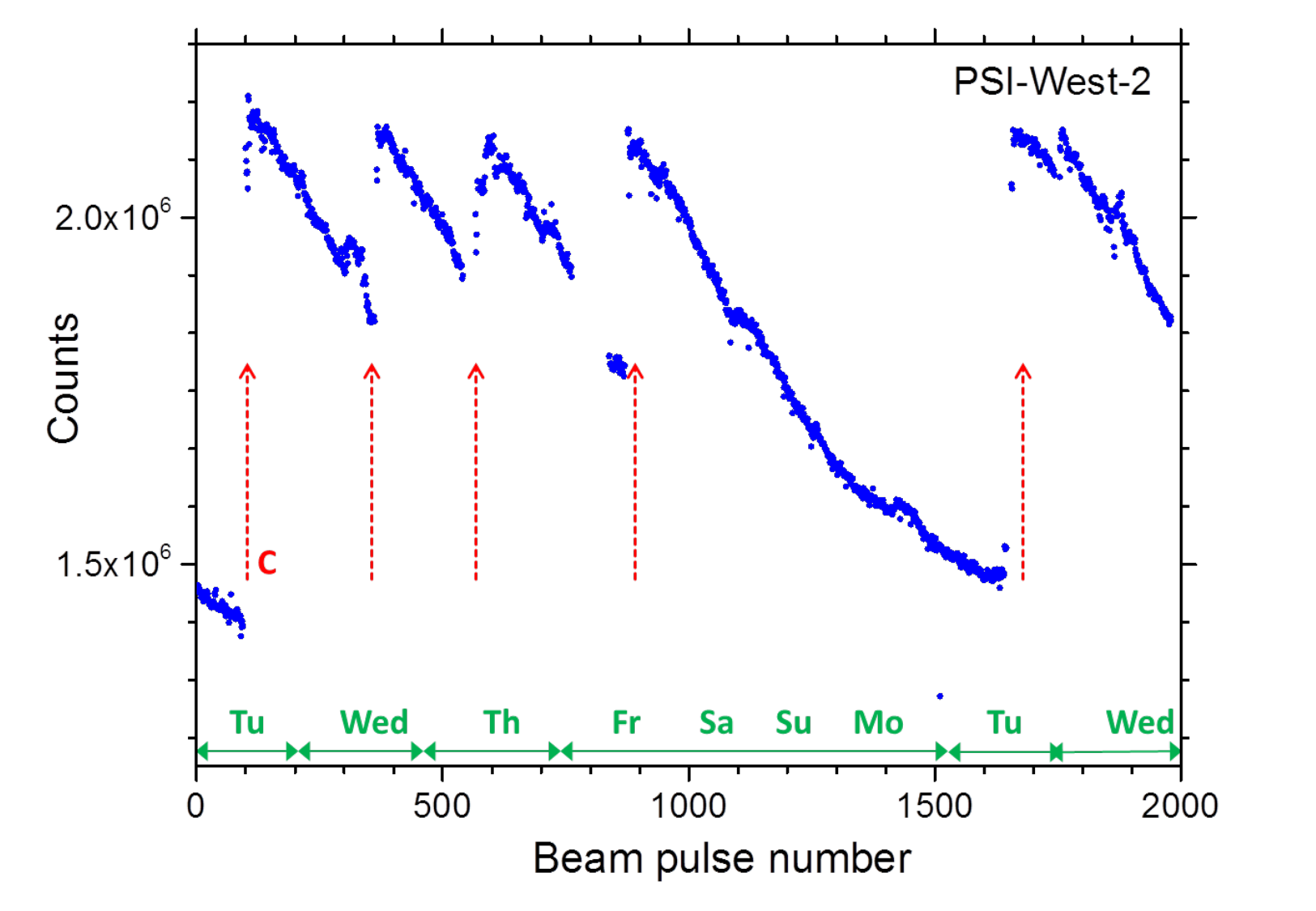}}
\caption{Decline and recovery of the UCN intensity 
as observed at the LANL (top) versus date,
and at the PSI (bottom) UCN source.
Days of the week are given for the PSI
measurement. The arrows indicate 
times where conditioning (C) procedures 
(see text) were performed.
}
\label{yield-decline-LANL-PSI}
\end{center}
\end{figure}

Meanwhile, three facilities are utilizing pulsed UCN production from sD$_2$. 
These are spallation sources
at LANL ~\cite{Saunders2004,LANL2000,LANL2013,Ito2018}
and PSI~\cite{Anghel2009,Lauss2011,Lauss2012,Lauss2014,Becker2015,Bison2017},
and at the Mainz TRIGA reactor source~\cite{Karch2014,Kahlenberg2017}. 
Although basic processes of UCN production and extraction from the source
are understood, there is a puzzling observation of a dynamic process
during operation of the pulsed UCN sources 
which is displayed in Fig.~\ref{yield-decline-LANL-PSI}.
The UCN yield in the sources at LANL and at PSI declines 
with the operating time of the sD$_2$, apparently correlated to the delivered proton beam charge,
at a rate of 5-20\% per day as shown 
in 
Fig.~\ref{yield-decline-LANL-PSI}\footnote{A decline was initially also observed at the 
Mainz TRIGA UCN facility~\cite{Karch2014}.
This decrease appears to either not be present or to
be masked by other effects 
in newer measurements~\cite{Kahlenberg2017}.}.
The initial UCN yield can be recovered as shown in the sequences 
of Fig.~\ref{yield-decline-LANL-PSI}. 
At LANL, 
the routine procedure is to melt and re-freeze the entire sD$_2$ volume 
of 1\,dm$^3$ about once a day.  
At PSI, the sD$_2$ volume is about 30\,dm$^3$.
Melting and refreezing is a long process for the PSI source and does not immediately 
improve the UCN yield. 
However, a procedure (``conditioning'') of treating the sD$_2$ for 
about 1.5 hours, leads to a full recovery of the UCN yield.
Conditioning includes two main steps, 
reduction of the D$_2$ cooling and 
powering the heaters mounted to the lid of the sD$_2$ vessel. 
Neither of these steps alone is sufficient to recover the UCN yield. 

In this paper we report
measurements of the UCN intensity decrease
observed at accelerator-driven sD$_2$ based UCN sources 
under continuous operation.
Observations at the PSI UCN source, discussed in detail in the next section,
suggest that the deterioration of the UCN yield is due to
surface effects. 
All indications point to the growth
of frost layers on the top surface 
which subsequently impede the transmission of UCN.
These findings are supported by the visual offline inspection 
of sD$_2$ surfaces under a heat pulse 
operation at the 
North Caroline State University (NSCU)
UCN facility~\cite{Korobkina2007}.
The UCN transport simulation in Sec.~\ref{simulation-psi}
strongly supports this interpretation of the observed UCN intensity decrease.

\subsection{Properties of cryogenic deuterium}

In order to make the following discussions concerning solid D$_2$ properties 
most accessible to the reader, 
we here recall some properties of deuterium, in particular of cryogenic D$_2$. 
For further reading, we refer the reader to the excellent, 
comprehensive book by Souers on the properties of the hydrogen isotopes~\cite{Souers1986} 
from which much of the background information and many of the values in this subsection are quoted. 

To guarantee good UCN performance, the UCN converter should consist of pure 
solid ortho-D$_2$ with as little as possible para-D$_2$ and H contamination~\cite{Morris2002}. 
While UCN losses due to upscattering and absorption are an important aspect of it, 
another one is the transparency of the sD$_2$ for UCN. 
Under all reasonable conditions for UCN sources, ortho-D$_2$ solidifies in 
hexagonal closed-packed (hcp) lattices. 
It is possible to grow optically transparent crystals from high purity ortho-D$_2$, 
however, the para-D$_2$ and H contaminations behave as crystal defects and 
make obtaining ``nice'' solids much harder. 
It is however known, that elastic UCN scattering on crystal defects can hinder UCN extraction from the solid into vacuum and thereby deteriorate UCN source performance~\cite{Atchison2005b}. 
Therefore, starting with D$_2$ of high isotopic purity  (99.8\% or, even better, 99.95\%) 
is mandatory as is the efficient conversion of para- to ortho-D$_2$. 
At room temperature, the equilibrium spin population of the D$_2$ molecules corresponds 
to the high-temperature limit of one third para-D$_2$ 
(there are in total three para-D$_2$, nuclear spin $I=1$ and six ortho-D$_2$, $I=0, 2$ sub-states). 
At low temperatures, the equilibrium goes to the lower energy ortho states, however, the natural transition rates are extremely slow. 
Therefore, UCN sources use catalyzed conversion to ortho-D$_2$, most commonly in the liquid state near the triple point on a paramagnetic substance, see~\cite{Liu2000,Bodek2004}. 
This guarantees high mobility of the D$_2$ molecules to reach the surface of the catalyst and as low temperature as possible. 

At the triple point, the D$_2$ temperature is 18.7\,K and its vapor pressure 171\,hPa. 
The equilibrium ortho concentration at the triple point is 98.5\%. 
This can be further enhanced as the UCN sources operate at lower temperatures. While the natural conversion remains slow, the rates can be considerably enhanced under irradiation~\cite{Collins1991}. 

An additional benefit of a high ortho-D$_2$ concentration is the high thermal conductivity of the solid, up to more than 30\,W/m/K at 5\,K but dropping quickly to around 1\,W/m/K at 12\,K. 
The heat capacity is strongly increasing with increasing temperature, 
starting from 0.5\,J/mol/K at 4\,K it is about 1\,J/mol/K at 8\,K and reaches 4\,J/mol/K at 12\,K.

In the UCN sources at LANL and at PSI, sD$_2$ is routinely frozen from the liquid phase, cooling more or less slowly through the triple point. 
The D$_2$ molar volume changes from 23.17 to 20.44\,cm$^3$/mol above and below the triple point, respectively, 
when going from liquid to solid. 
The contraction of the solid D$_2$ continues when further cooling down, 
the molar volumes are 20.21, 19.97, 19.93\,cm$^3$/mol at 16, 10, 4\,K, respectively. 
It is known that thermal stress induced crystal defects can considerably increase the UCN elastic scattering in the solid. 
These can occur when cooling down to operational conditions but also in thermal cycles when operating the UCN source in a pulsed mode~\cite{Atchison2005b}.

The saturated vapor pressure of the solid depends  
strongly on temperature with 
pressures of 3390, 98, 6.5, 0.12, and 0.0002\,Pa at 16, 12, 10, 8, and 6\,K, respectively. 
In principle it is possible to measure the equilibrium temperature of the sD$_2$ surface by measuring the D$_2$ vapor pressure. 
In the pressure data of the PSI UCN source shown below, the situation is more complicated.  
The pressure gauges are limited and cannot measure pressures below a fraction of a Pa. 
Also, one must deal with transient pressures during pulsed operation measured with gauges at the end of roughly 10\,m long gas pipes away from the sD$_2$.
Additionally, a small contamination of the D$_2$ gas with $^4$He prevents the straight forward 
interpretation, however, allows for a more complicated analysis 
using the $^4$He as a gas thermometer. 
This analysis is much more involved and beyond the scope of the present paper.


\section{Experimental observations}

\subsection{Measurements at the PSI facility}
\label{Sec:PSI}


The UCN source at PSI is based on spallation of lead 
using the full 1.3\,MW, 590\,MeV proton beam~\cite{cyclotron2010}
directed
onto the UCN spallation target~\cite{Wohlmuther2006} for up to 8\,s.
Spallation neutrons are subsequently thermalized in heavy water,
a process that is well understood and modeled in a 
full MCNP-X simulation~\cite{Becker2015}
containing all important geometry details of the target 
and moderator area of the UCN source.
Thermal neutrons can then enter the moderator vessel
cooled to 5\,K 
containing about 30\,dm$^3$
of solid ortho-deuterium,
where 
they are moderated to cold neutrons. 
Finally, they may become ultracold via phonon 
down-scattering in the 
sD$_2$~\cite{Golub1991,Atchison2007,Atchison2011,LANL2000,Kirch2010,Atchison2009b,Serebrov1997,Young2014}.
During the proton pulse UCN are produced and then stored in a
2\/m$^3$ large storage vessel, providing UCN to three beamlines.
In recent years a detailed model
of the UCN transport in the PSI UCN source
has been developed.
Typical pulse repetition intervals
are 300\,s or 500\,s.
To obtain information on the source performance, 
UCN can be counted at a beamport, integrating 
over one pulse interval; 
the corresponding counts are shown 
in Fig.~\ref{yield-decline-LANL-PSI}-bottom for a one week period.
The intensity decrease with time is obvious.
Full restoration of the UCN intensity is achieved in
a conditioning procedure (see below).

\begin{figure}[htb] 
\begin{center}
\resizebox{0.5\textwidth}{!}{\includegraphics{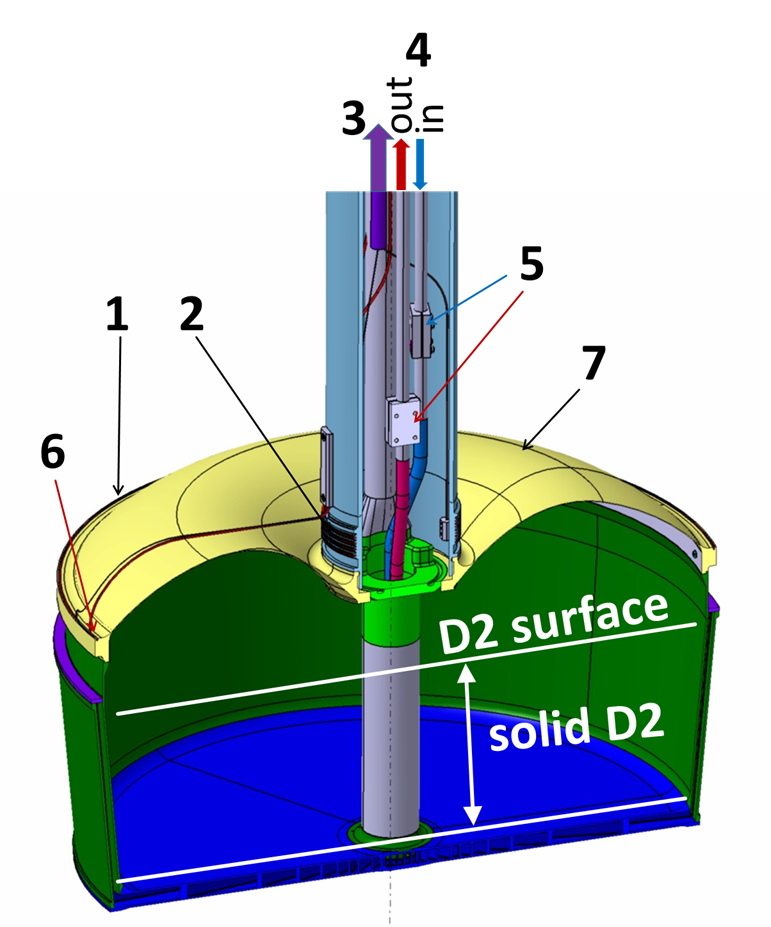}
}
\caption{Cut view of the PSI sD$_2$ vessel
with inside diameter of 48\,cm at the D$_2$ surface.
1: outer lid heater (15\,W);
2: inner lid heater (15\,W);
3: line towards pressure sensor (CP006);
4: He cooling line in and out;
5: He temperature sensor on the inlet (CT031) and outlet (CT032);
6: thermocouple (CT302);
7: vessel lid (AlMg3).
A photo can be found in Ref.~\cite{Lauss2012}. 
}
\label{moderator-vessel}
\end{center}
\end{figure}

A cut view of the moderator vessel ($\invdiameter$=48\,cm)
made mainly from AlMg3 is shown in 
Fig.~\ref{moderator-vessel}
with indicated approximate height of the sD$_2$
of about 15\,cm maximum level.
The vessel is cooled by supercritical helium 
at 4.5\,K 
which enters through a central pipe
and flows radially through the bottom plate and up the side walls
and back. 
The cooling pipes are indicated in Fig.~\ref{moderator-vessel}.
The vessel is equipped with 
temperature sensors 
for the He lines (Cernox)
and 
on the vessel lid (thermocouple Type E - sensor CT302).
The gas pressure above the deuterium is measured 
about 10\,m away at room temperature
with a gas-type independent
capacitance diaphragm gauge (CP006)
connected via a DN25 tube.

\begin{figure}[htb] 
\begin{center}
\resizebox{0.5\textwidth}{!}{\includegraphics{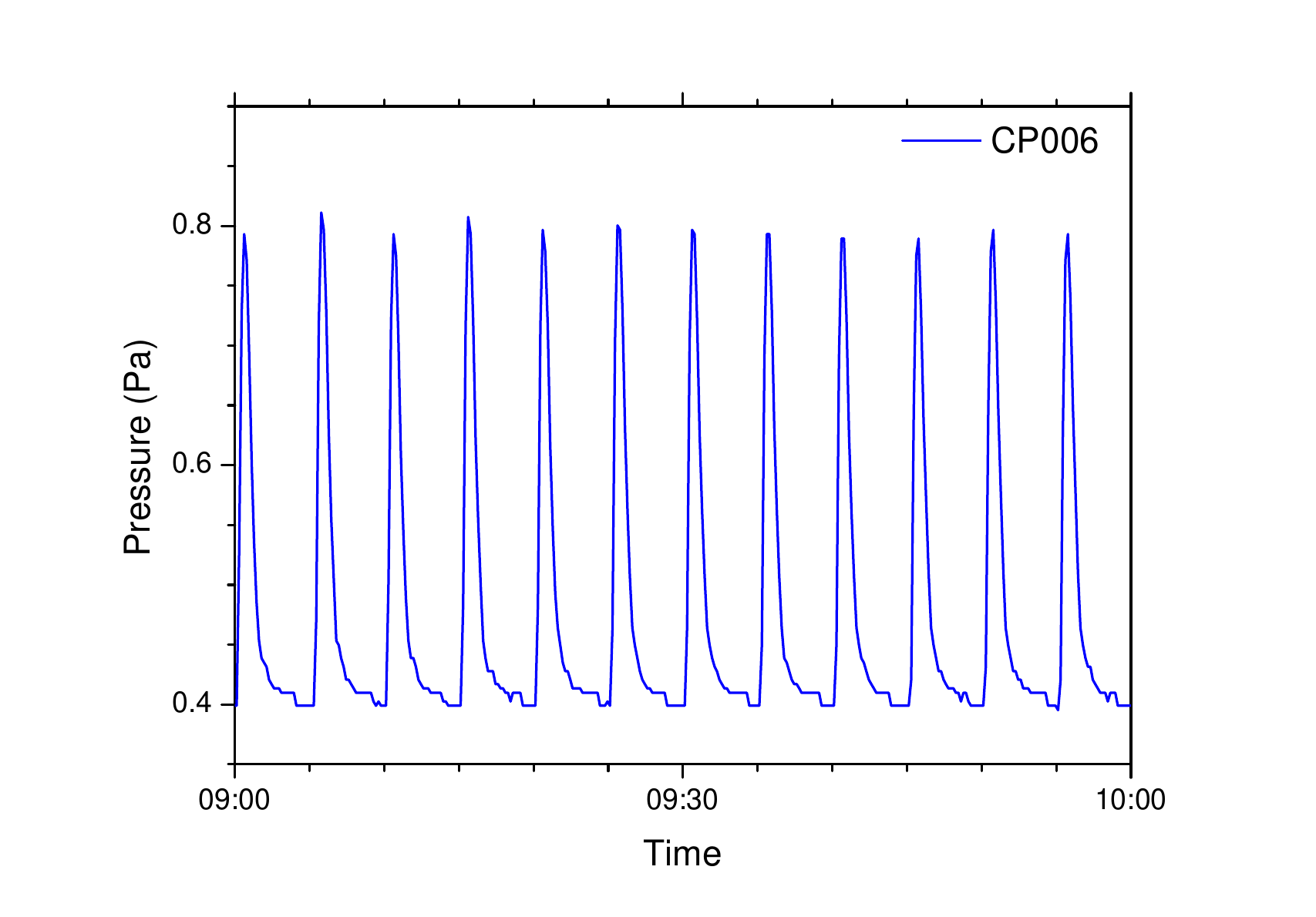}
}
\resizebox{0.5\textwidth}{!}{\includegraphics{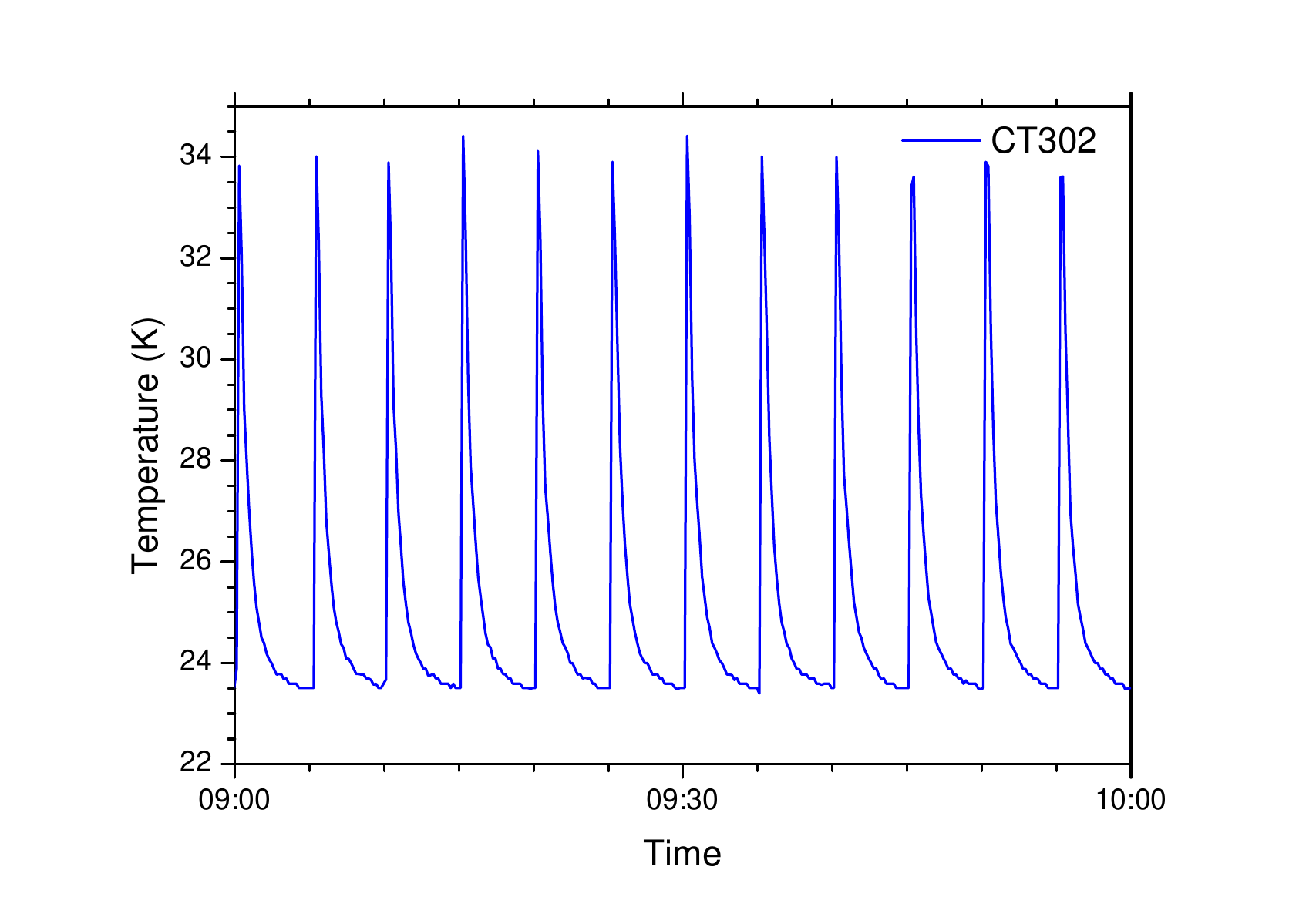}}
\caption{Measurements at PSI during a period of 8\,s long beam pulses, 
pulse repetition interval of 500\,s.
Top: Pressure behavior above the sD$_2$ surface (sensor CP006);
bottom: Temperature behavior on the vessel lid (sensor CT302).
Both parameters rise during the time of the proton beam impinging on the target 
and decrease
to the equilibrium value before the next pulse.
The baseline pressure of 0.4\,Pa
is due to residual $^4$He in the D$_2$ gas system 
and does not reflect
the much lower 
D$_2$ saturation vapor pressure.
}
\label{pressure-PSI}
\end{center}
\end{figure}

In order to assess the impact on the sD$_2$ of beam pulsing,
we show in Fig.~\ref{pressure-PSI}
the pressure above the sD$_2$ 
and the temperature
at the outside rim of the vessel lid
for a period of continuous pulse operation.
Both sensors show a steep increase 
in pressure and temperature during the proton
beam pulse and a relaxation afterwards towards 
the equilibrium value. 

\begin{figure}[htb]
\begin{center}
\resizebox{0.5\textwidth}{!}{\includegraphics{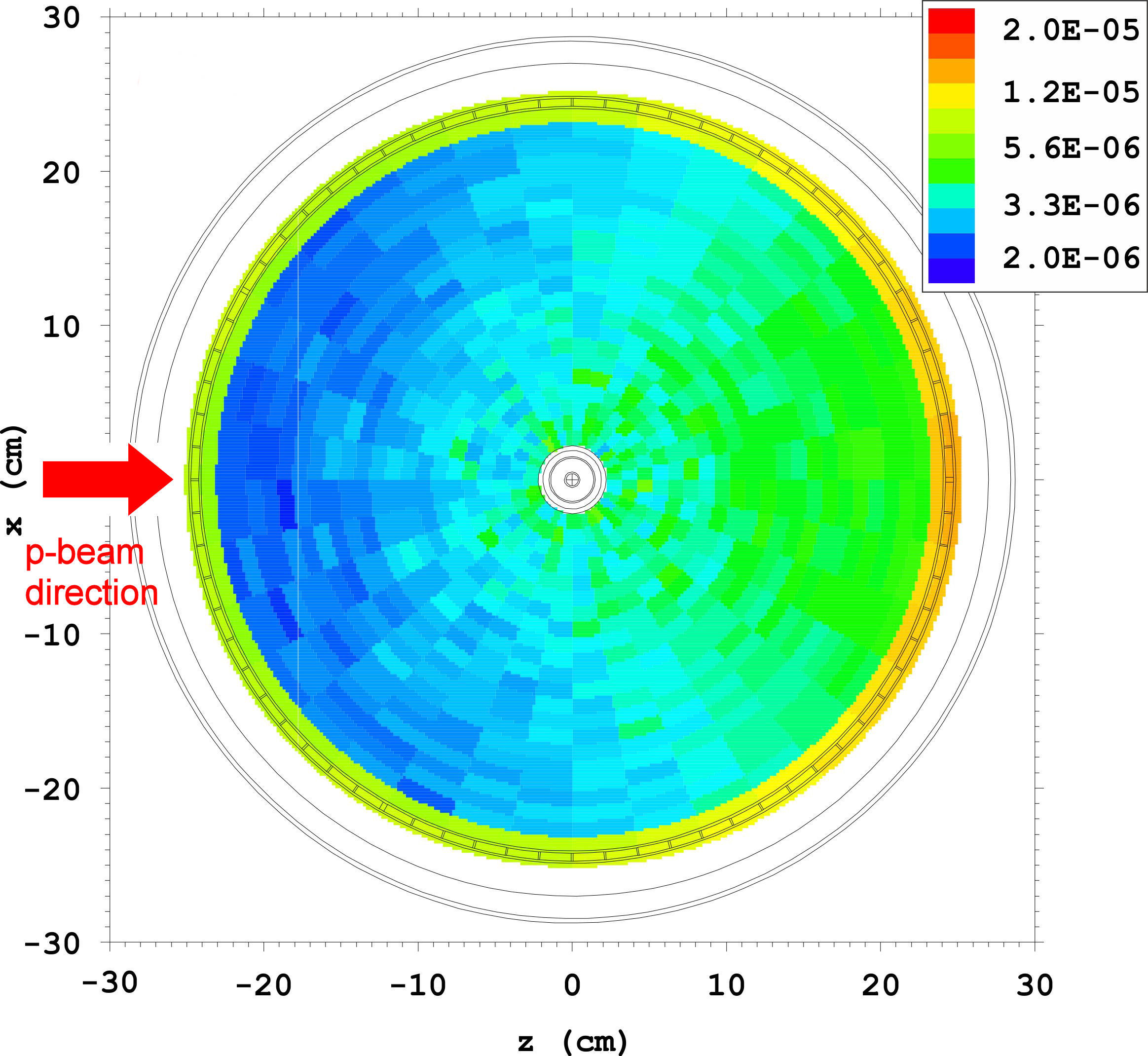}}
\caption{MCNP-X simulation of the deposited
energy density (MeV/cm$^3$ per proton onto the target)
in the top 1\,cm layer of the sD$_2$
and in the Al of the moderator vessel.
The proton beam direction is indicated.
}
\label{PSI-energy-density}
\end{center}
\end{figure}

MCNP-X is also used to simulate the
energy deposition in the relevant source materials~\cite{Talanov2017}.
Fig.~\ref{PSI-energy-density}
shows the energy density (MeV/cm$^3$) per proton on target
deposited in the 
top 1\,cm of the sD$_2$ bulk.
The total of $\gamma$-ray and neutron heating
for an 8\,s pulse of 2.2\,mA results
in an energy deposition of
about 0.1\,J/cm$^3$ in the sD$_2$
in that layer.
The asymmetry 
is caused by the asymmetry in the neutron flux
reflecting the proton beam direction (origination at -z).
Higher power deposition occurs in the aluminum material
of the vessel, which shows up as the colored outside ring.

The UCN intensity at PSI, plotted in 
Fig.~\ref{yield-decline-LANL-PSI}-bottom, 
shows that periods of deterioration 
over 1 day are followed by a short period without
beam, where conditioning is done.
This results in a reliable full recovery of the output intensity.
The conditioning procedure consists of two components:\newline
Part 1: Reducing 
the standard 
He coolant flow to a minimum value.
The flow is not completely stopped in order to prevent
instabilities in the cooling circuit when returning to regular cooling conditions.\newline
Part 2: Switching on the lid heaters located on the moderator vessel
as shown in Fig.~\ref{moderator-vessel}.
Inner and outer lid heaters are turned on to a
power of 7.5\,W each.

These steps together cause a temperature increase 
of the D$_2$ surface region 
to 10-12\,K as derived from the vapor pressure.
After about 60\,min 
the cooling power is gradually restored over 30 minutes
to regular conditions.
Finally both heaters are switched off.
It is important to note that only the combination 
of cooling flow reduction and lid heating ensures the
full recovery of the UCN yield.

\begin{figure}[htb]
\begin{center}
\resizebox{0.5\textwidth}{!}{\includegraphics{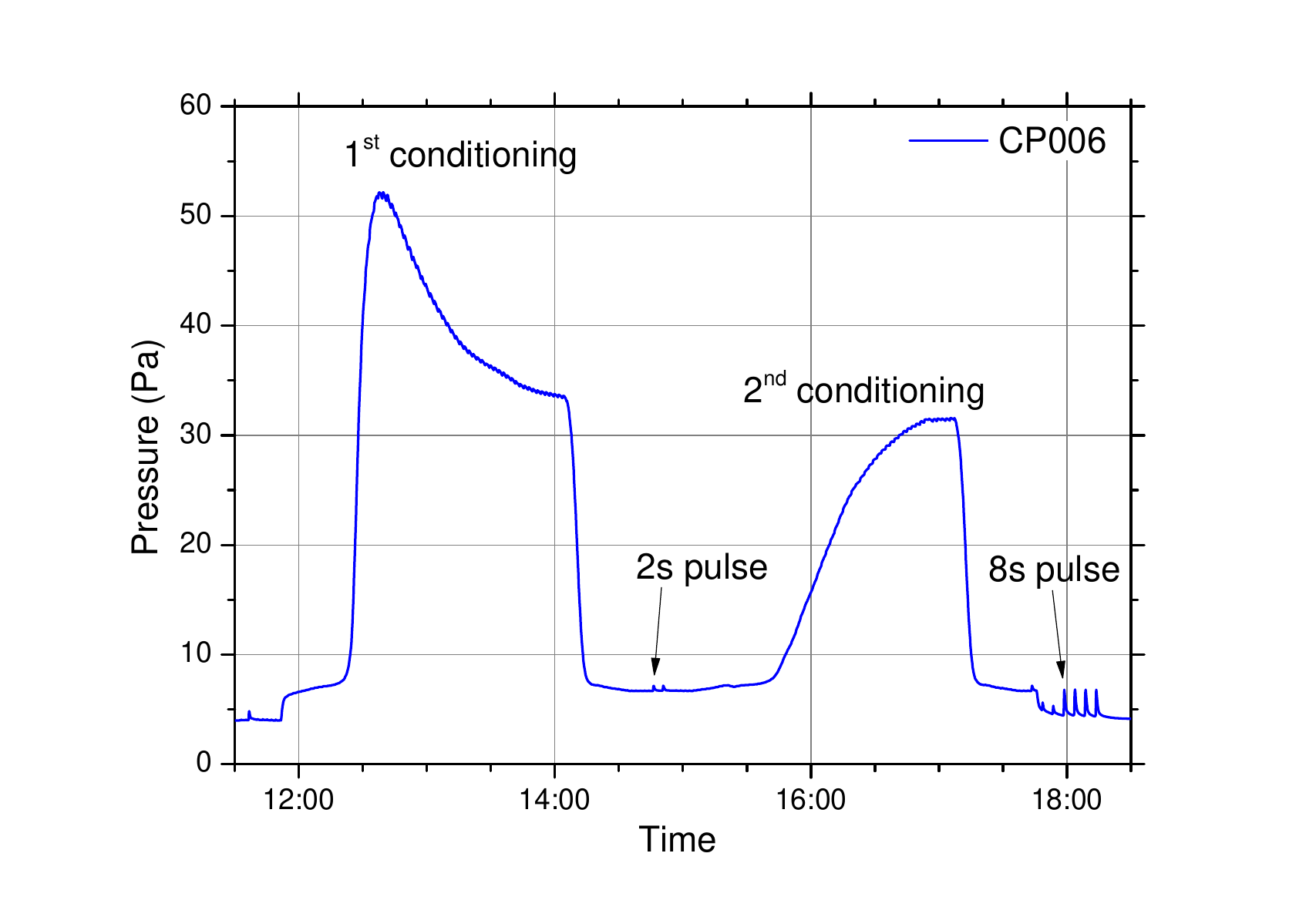}}
\caption{
D$_2$ vapor pressure
showing a double conditioning procedure as explained in the text.
The standard single conditioning procedure 
is shorter than 
the first conditioning 
shown here (period ~12:00 - 14:00).
One can see that the pressure after reaching its maximum
decreases again without any changes in cooling to a level
which corresponds to the the pressure level reached in the
subsequent conditioning procedure.
The minimum pressure between the two conditioning periods
is a bit higher than before and after, as the lid heating 
was not switched off.
}
\label{PSI-conditioning-2}
\end{center}
\end{figure}

Figure~\ref{PSI-conditioning-2} shows the evolution of
the D$_2$ vapor pressure (sensor CP006) during 
two subsequent conditioning
procedures.
In procedure 1,
following a UCN source operation period with 
1112 pulses,
the vapor pressure rises to a maximum of around 53\,Pa
and then,
without any change to the cooling conditions,
drops to around 33\,Pa, similar to
the maximum pressure level observed in the second procedure
which was performed subsequently.
Between the 2 procedures 
two 2\,s long proton pulses were used 
to check that the 
UCN yield was fully regained.
The second conditioning resulted in no further 
gain of the UCN yield.
During regular source operation the 
total time used for a conditioning period can
be as short as 90\,min with
full UCN yield recovery.

During regular UCN source 
operations (see Fig.~\ref{yield-decline-LANL-PSI}, bottom,
single conditioning procedures were
performed many times over several months and
the same pressure behavior 
was observed.

We interpret these pressure curves as follows:
During every proton pulse some D$_2$ sublimates, and
a small amount of it contributes to the buildup of
a layer that is poorly attached to the sD$_2$ surface,
a frost/snow-like layer.
After operating the UCN source for an extended period with 
several tens to hundreds of 
proton pulses a conditioning procedure is performed.
During this conditioning the D$_2$ surface warms up 
and sublimation of D$_2$ occurs
first from the frost layer.

Sublimation of this layer causes the steep pressure increase
up to a maximum value, 
e.g., around 53\,Pa 
as in Fig.~\ref{PSI-conditioning-2}.

After a certain time, most of the badly-cooled D$_2$ frost is sublimated 
and resublimated (presumably as better quality D$_2$ ice)
at colder surfaces causing the pressure to decrease despite the reduced cooling.
A lower pressure level is then established towards the end 
of the first conditioning, which corresponds to the maximum level obtained 
in the subsequent second conditioning of Fig.~\ref{PSI-conditioning-2}.
The two, short 2\,s, long proton beam pulses applied for UCN gain monitoring 
between the conditioning procedures 
do not produce appreciably new frost. 

\begin{figure}[t]
\begin{center}
\resizebox{0.5\textwidth}{!}{\includegraphics{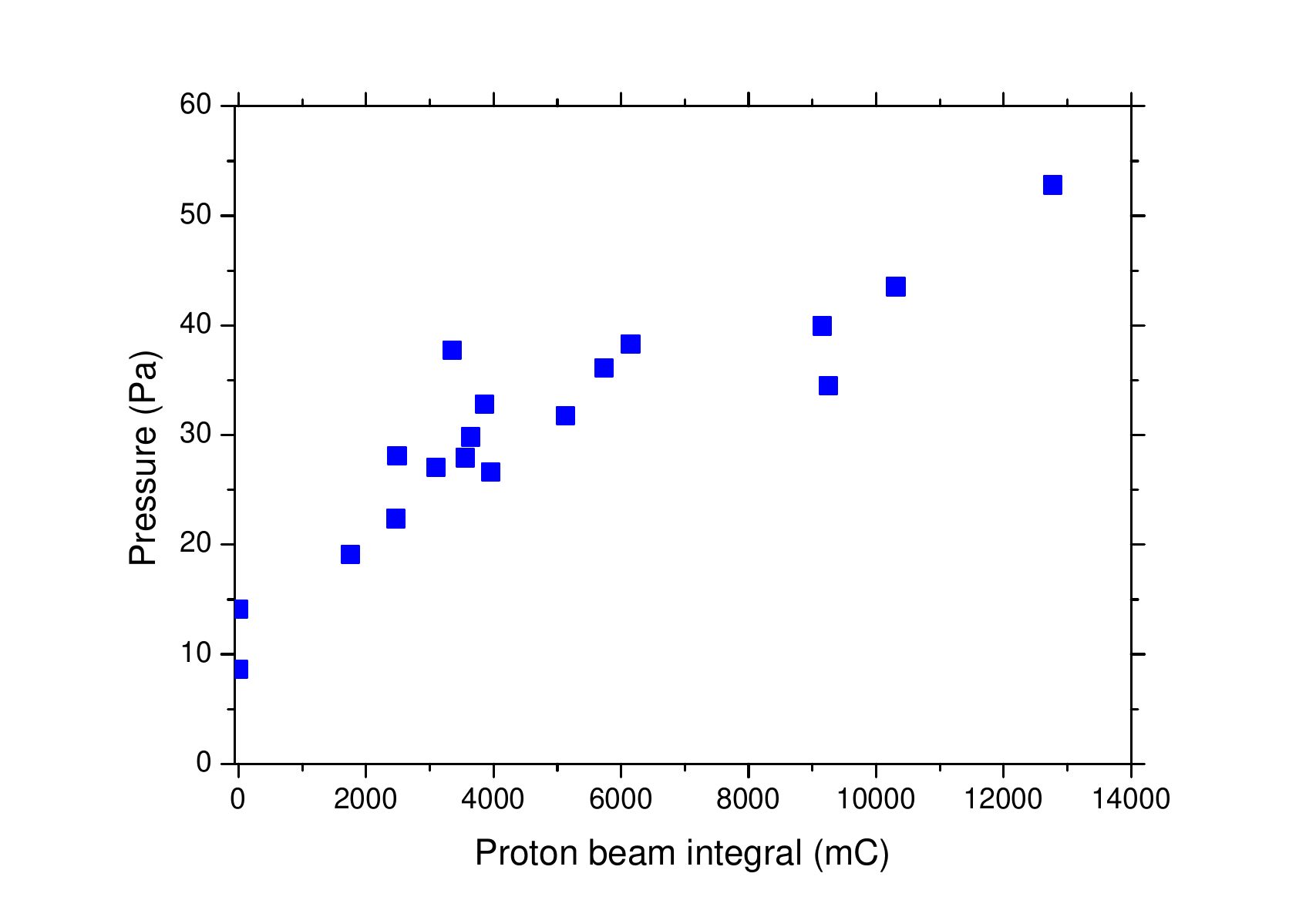}}
\caption{
Correlation between the maximum 
pressure during conditioning 
and the integrated proton beam current 
in the time period since the previous conditioning.
}
\label{PSI-pressure-vs-proton}
\end{center}
\end{figure}

The build up of frost layers is proportional to the accumulated 
(total) number of pulses.
Fig.~\ref{PSI-pressure-vs-proton} displays data from standard conditioning procedures 
performed in an operating period in fall 2017.  
One observes a clear correlation between the maximum 
vapor pressure in the conditioning 
procedure -- corresponding to more material evaporated -- and the number of beam 
pulses, i.e. the total proton charge since the previous conditioning.
The growing layer thickness results in a continuous reduction of the UCN yield 
as observed and supported by UCN simulations, see Sec.~\ref{simulation-psi} below.

Measurements of the UCN intensity
over time were performed 
at beamports West-1 and West-2 at the PSI UCN facility 
detailed in Fig.1 of Ref.~\cite{Lauss2012}, where a 
Cascade UCN detector\footnote{C-DT, Hans-Bunte Str. 8-10, Heidelberg, Germany}
was installed directly at the beamport.
The West-1 beamport is located about 1.13\,m 
and the West-2 beamport 3.23\,m 
above the sD$_2$ surface.
This results in a relative shift of the 
UCN energy spectra at the two beamports.
UCN which can reach the higher beamport have a minimum energy of 320\,neV
outside the sD$_2$ surface, while UCN at West-1 have a minimum energy
of 113\,neV, similar to the kinetic energy boost by the sD$_2$ 
potential~\cite{Daum2008,Altarev2008}. 
Both beamlines are equipped with identically produced UCN guides made from glass tubes
with inside coating of nickel-molybdenum~\cite{Blau2016}.

In 2016 and 2017, the UCN yield decrease 
during operation
was observed
over months of continuous UCN source operation,
i.e. when pulses typically of 8\,s length and with 300\,s repetition interval
were delivered continuously.
These were the operating conditions for the nEDM experiment 
at PSI~\cite{Baker2011,Pendlebury2015}.
No UCN yield change was observed 
when the sD$_2$ was left undisturbed, 
even after several days without pulses.

\begin{figure}[htb] 
\begin{center}
\resizebox{0.5\textwidth}{!}{\includegraphics{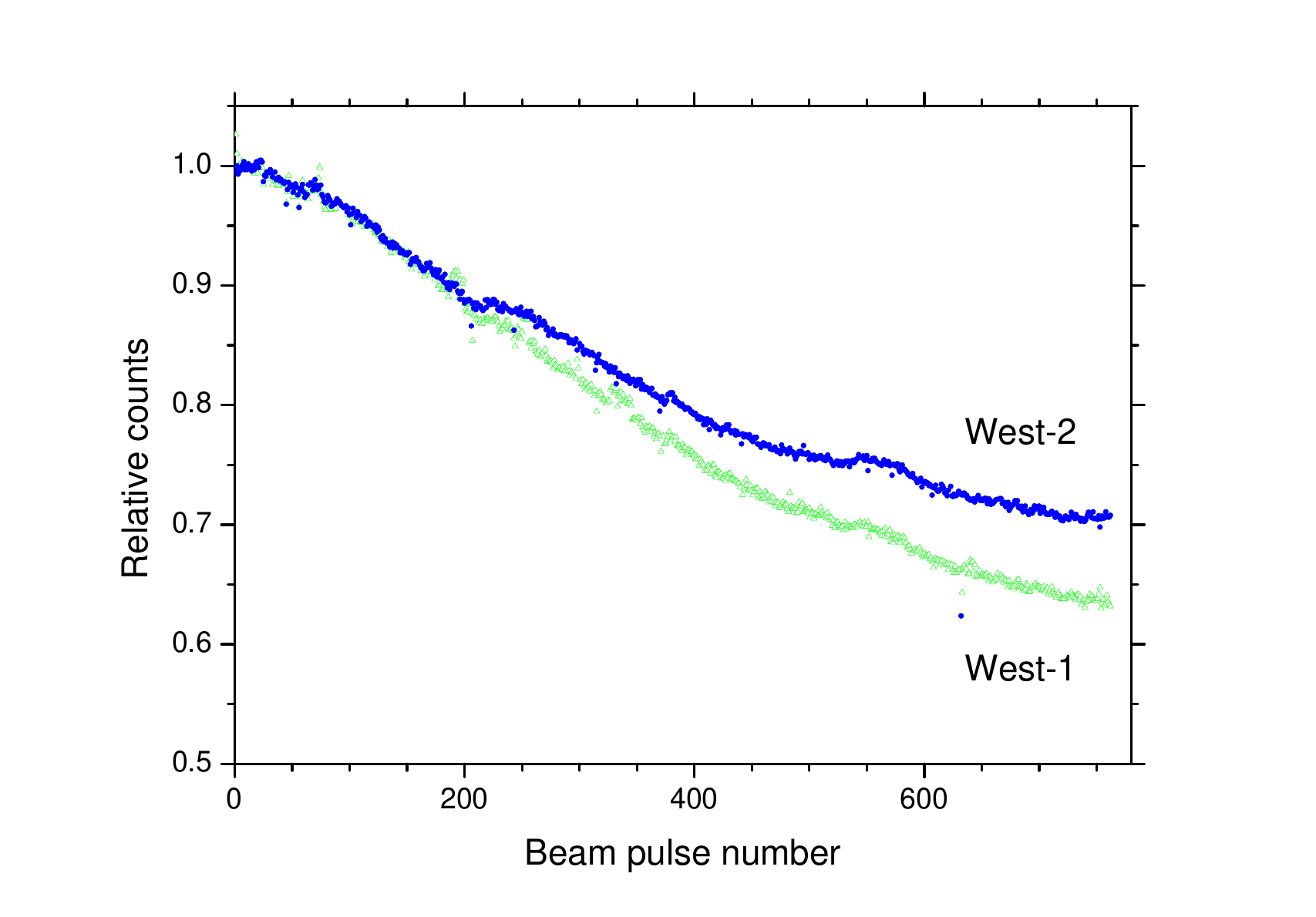}
}
\caption{
Relative UCN counts versus proton beam pulse number
showing the decline of the UCN intensity over a three days period
with pulses every 300\,s,
as observed on beamport West-1 and West-2.
The small correlated intensity variations are caused by proton beam
fluctuations.
}
\label{yield-decline-PSI-weekend}
\end{center}
\end{figure}

\begin{figure}[htb]
\begin{center}
\resizebox{0.5\textwidth}{!}{\includegraphics{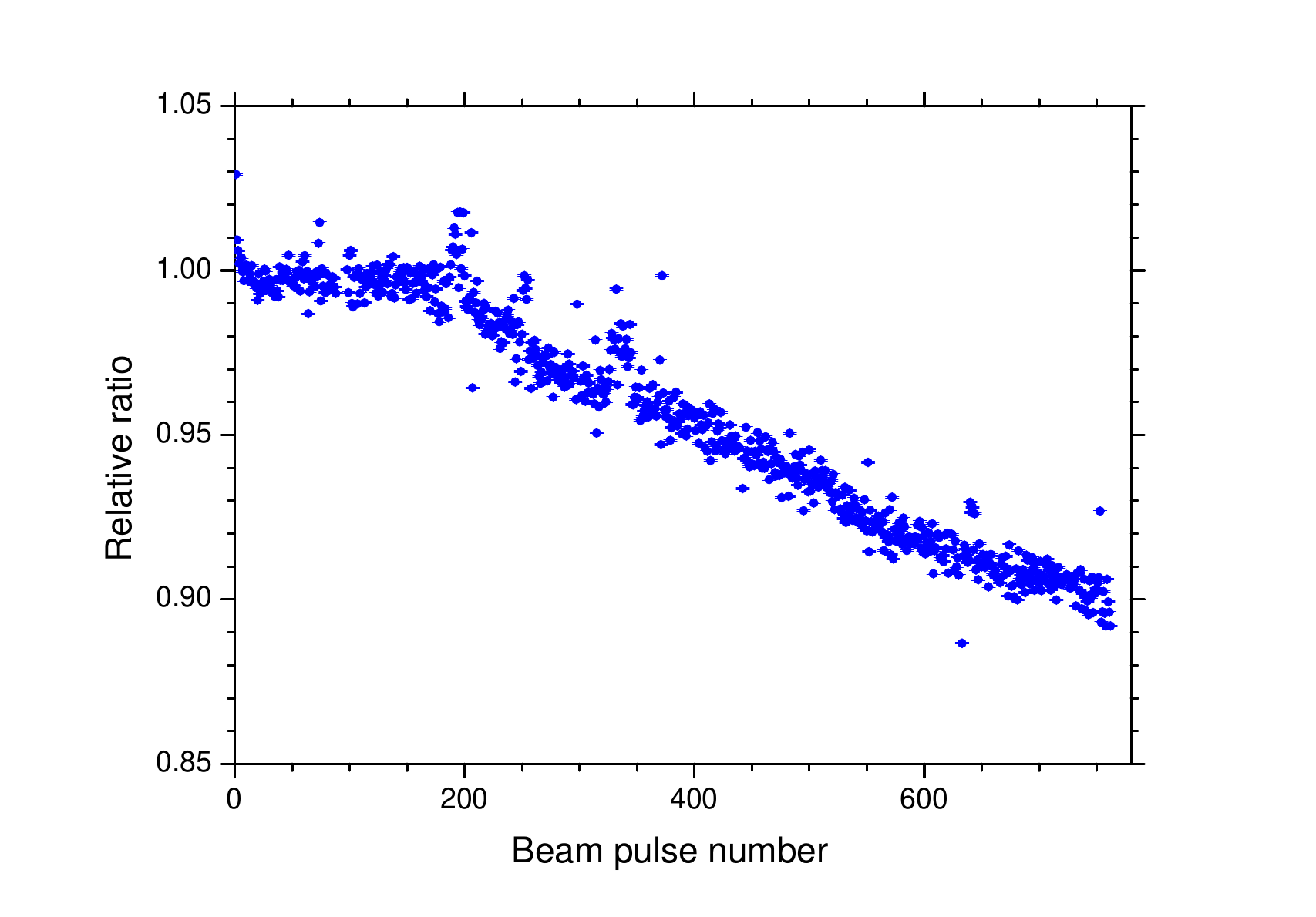}
}
\caption{
Ratio of relative counts in West-1 
and West-2
for the period shown in Fig.~\ref{yield-decline-PSI-weekend}.
}
\label{yield-ratio-PSI}
\end{center}
\end{figure}

The long period of decrease, observed  
in Fig.~\ref{yield-decline-LANL-PSI}, corresponds to three days
with continuous pulse operation without conditioning.
Figure~\ref{yield-decline-PSI-weekend}    
compares the 
UCN intensities for West-1 and West-2 during 
this period.
Correlated fluctuations are due to variations 
in proton beam current.
A clear difference is observed at the two beamports, 
with the rate decline
being more pronounced in West-1.
Figure~\ref{yield-ratio-PSI}   
shows the ratio of relative counts observed 
at the two beamports (West-1/West-2).
After the conditioning procedure, 
the ratio stays constant for a few hours
but
then starts to decrease continuously. 
This decrease can be attributed to the formation of a
growing frost layer on top of the D$_2$ surface. The UCN
rate at the lower beamport decreases faster than at the
higher beamport since only UCN passing the surface with higher
kinetic energies are able to reach the top beamport, and these
are less affected by a possibly energy dependent transmission loss
in the frost layer. 
The initial constant behavior is the subject of
ongoing research.


\subsection{UCN measurements at LANL UCN source}

\begin{figure}[htb]
\begin{center}
\resizebox{0.5\textwidth}{!}{\includegraphics{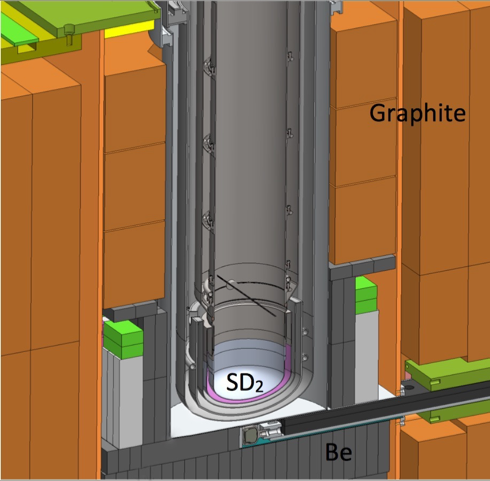}
} 
\caption{Geometry of the most recent upgrade of the LANL UCN source.}
\label{fig:LANL_geometry}
\end{center}
\end{figure}

Solid deuterium UCN source development at LANL began with a series of 
prototypes~\cite{LANL2000}, 
followed by commissioning the first ``production'' source in 2005, 
which ran until 2016~\cite{LANL2013} when a major upgrade 
to the source was implemented \cite{Ito2018}.  
The upgraded source is currently in operation.  
The later prototypes and subsequent sources all have very similar geometries, 
consisting of a sD$_2$ source situated at the bottom of 
a roughly 1\,m vertical guide, which is coupled via 
an ``elbow'' at the top to horizontal guides which exit the biological shielding.  
A ``flapper'' valve located immediately above the sD$_2$ opens 
during the proton pulse and then shuts rapidly to 
prevent UCN loaded into the guide system from re-entering 
the sD$_2$ source between pulses.  
The most recent geometry of the source is shown 
in Fig.~\ref{fig:LANL_geometry}.  
The UCN yield decline is observed in both production sources.
As was mentioned in the introduction, 
the sD$_2$ volume is $\simeq$1\,dm$^3$, which can be easily 
melted and re-frozen to recover UCN intensity. 
The LANL source repair procedure is different from the PSI procedure because it
involves a full melt and re-freeze of the bulk sD$_2$, 
resulting in changes not only to the surface, 
but also to the bulk properties of the sD$_2$.
Note that while the source operators have attempted
to develop a reproducible procedure,
variations are unavoidable.
As will be explained in the simulation section, 
the total frost effect depends on the bulk properties, 
so the rate of UCN yield decline varies more for the LANL source than the PSI source.

\begin{figure}[htb]
\begin{center}
\resizebox{0.5\textwidth}{!}{\includegraphics{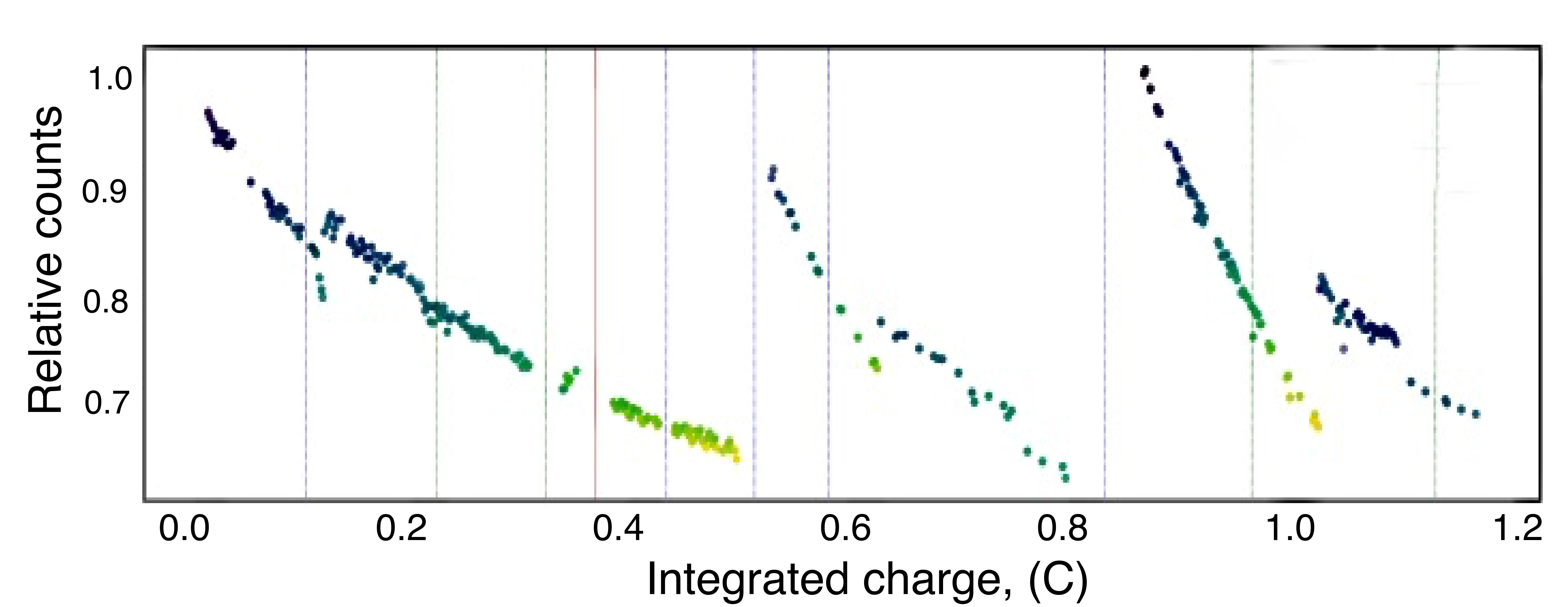}
} 
\caption{
Normalized counts of the LANL gate monitor versus integrated charge
for the same period as plotted in Fig.~\ref{yield-decline-LANL-PSI}-top.
The comparison shows that there is no UCN yield attenuation without the beam running.  
The color scheme gives an indication of spectral ``hardening'' with dark (blue) shading,
indicating a larger fraction of low energy neutrons seen in monitor detectors.
(see Fig.~\ref{fig:Beam-Tau-Ratio}).
}
\label{fig:LANL-charge}
\end{center}
\end{figure}

The LANL UCN source has two sets of detectors monitoring the UCN rate, 
located at different elevations, and therefore, probing different parts of the UCN spectrum. 
The first location is near the main gate valve just outside the
biological shield, which separates the UCN source from the neutron guide network, 
connected to different experiments~\cite{LANL2013}. 
There are two detectors at this location, called upstream normalization detectors,
probing the total UCN spectrum coming from the source. The count rate of one of these
monitors is shown in Fig.~\ref{yield-decline-LANL-PSI} versus date and
in Fig.~\ref{fig:LANL-charge} versus integrated beam charge.
The comparison shows that there is no UCN yield attenuation without the beam running.  
The second set of detectors is part of the UCNTau magnetic storage 
experiment~\cite{Young2014,Pattie2017}.
%
This set, called the elevated normalization detectors, 
has the highest detector elevated about 1\,m above the main beam height.

\begin{figure}[htb]
\begin{center}
\resizebox{0.5\textwidth}{!}{\includegraphics{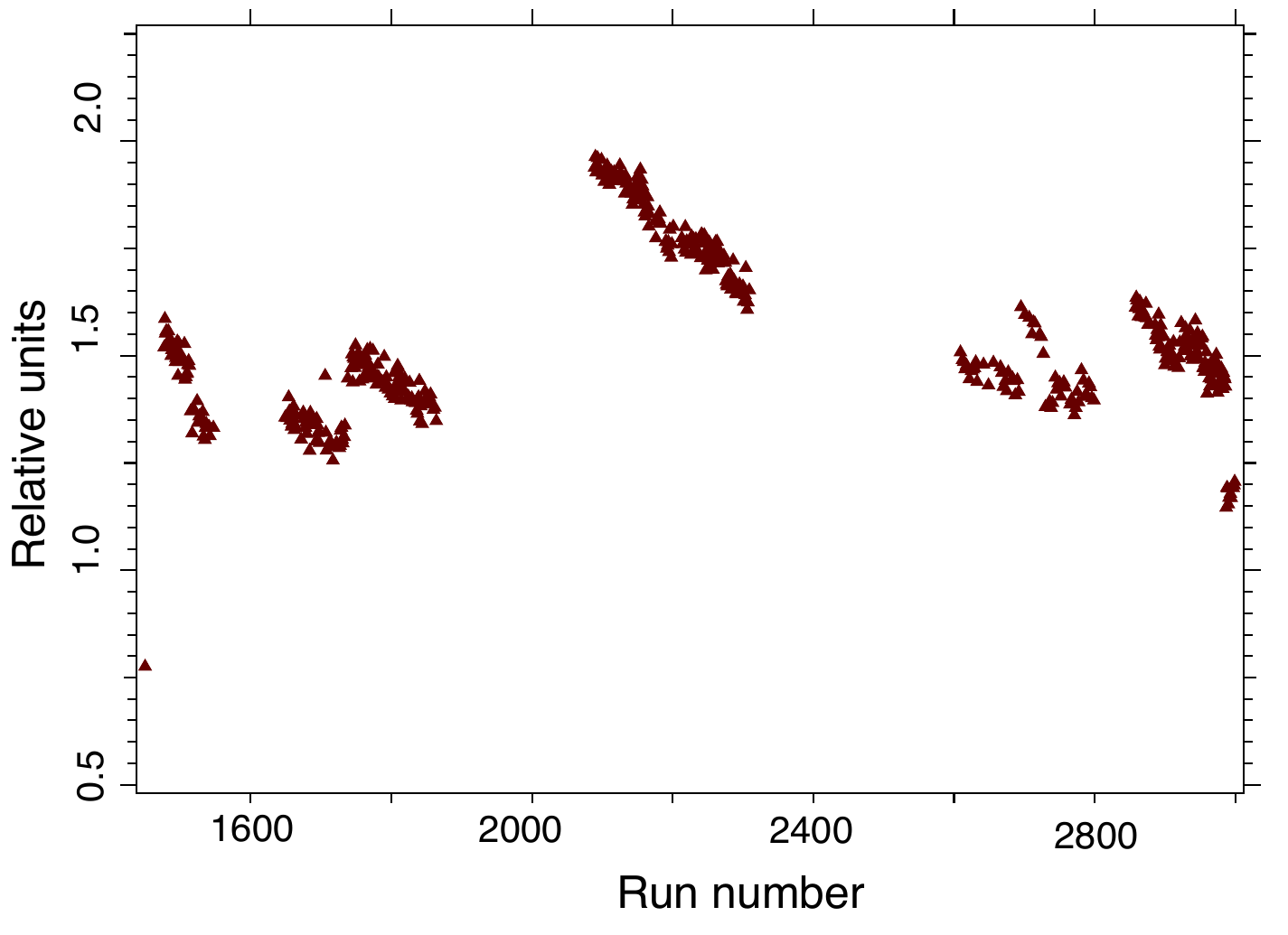}
} 
\caption{Ratio of total to high-energy UCN counts measured at the LANL UCN source.}
\label{fig:Beam-Tau-Ratio}
\end{center}
\end{figure}

The ratio of UCN count rates of the upstream to the highest elevated monitor counter 
versus time is shown in Fig.~\ref{fig:Beam-Tau-Ratio}.  
Compared with a monitor
detector proportional to the total yield (Fig.~\ref{yield-decline-LANL-PSI}), 
one clearly can see the correlation
between the spectrum-sensitive ratio and the overall yield from the source.
The decrease of the ratio is due to relative increase of
the high-energy part of the UCN spectrum, the same effect as observed at PSI.  

\subsection{Optical observation of sD$_2$ at the NCSU UCN source}
\label{optic-NCSU}

Optical monitoring of the evolution of surface structures on a 1\,dm$^3$, 
sD$_2$ crystal was
performed at the PULSTAR reactor at NCSU,
using the cryostat and cooling system of the UCN source~\cite{Korobkina2007}, 
which has completed cryogenic commissioning but has not yet been
placed into the biological shield.  
Details of the UCN source design are published in Ref.~\cite{Korobkina2014}.

\begin{figure}[htb] 
\begin{center}
\resizebox{0.5\textwidth}{!}{\includegraphics{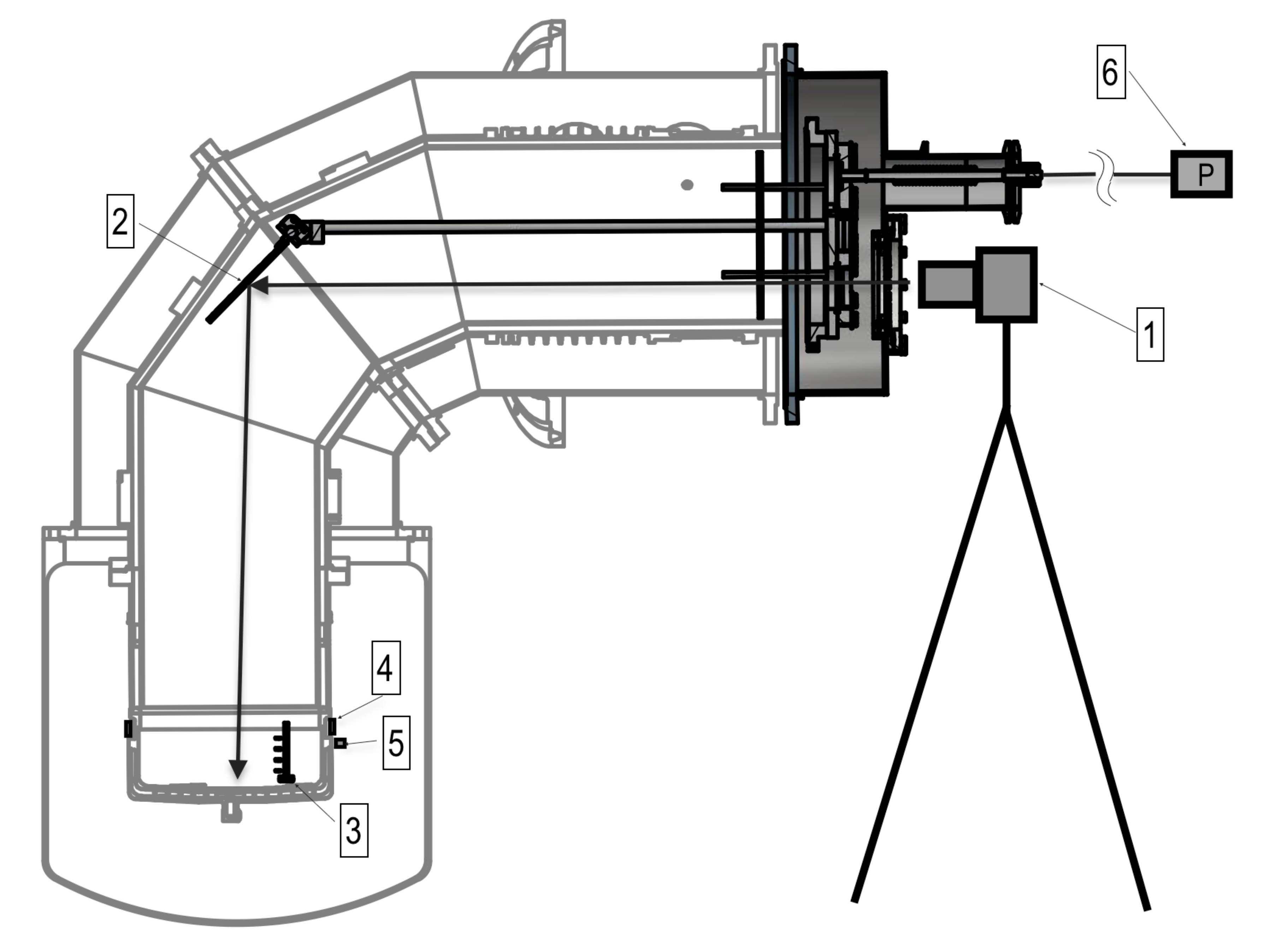}
}
\caption{Instrumentation for visual and temperature control 
of sD$_2$ growth in situ at the PULSTAR UCN source cryostat: 
1: camera; 
2: stainless steel mirror; 
3: holder with four temperature sensors; 
4: heater; 
5: temperature sensor; 
6: pressure sensor at about 10\,m distance, 
connected by a 12.5\,mm diameter tube.}
\label{NCSU-source}
\end{center}
\end{figure}

The geometry of the deuterium vessel (elbow) and 
deuterium volume at PULSTAR (1\,dm$^3$) 
is more similar to the geometry of the LANL source than the PSI source,
most importantly, because
there is no lid above the sD$_2$. 
This permits direct visual monitoring of the deuterium growth
using a special flange with instrumentation designed 
specifically for this test (see Fig.~\ref{NCSU-source})~\cite{Medlin2017}.

Pictures of the sD$_2$ inside the cryostat were taken 
with a camera on time lapse.  
Optical windows are mounted on both flanges, 
vacuum jacket, and cryostat insert.  
An additional optical window on the inner flange, 
above the camera window, 
was used to illuminate the deuterium 
from a light emitting diode. 
Since the cryostat has the shape of an elbow, 
a stainless steel mirror was installed on long rods.  
After mirror insertion, the angle was adjusted for the best view of the top of the sD$_2$ 
A holder with four diode temperature sensors,
mounted on the same rods, 
was lowered onto the bottom of the deuterium container 
to probe the bulk temperature of the sD$_2$
One additional temperature sensor was installed on the outside wall of the container, 
right below the heater. 
A 25\,W heater,
made of the aluminum alloy Al6061,
was clamped around the top of the sD$_2$ container.
Above the heater, aluminum walls were welded to a 3 inch long Zircalloy ring 
to thermally decouple the top part of the cryostat insert, 
also made of Al6061.
This part was held above 40\,K to prevent deuterium ice formation with the
D$_2$ triple point being at 18.7\,K. 
The bottom of the container was cooled by a cryogenic flow of helium 
injected through the cooling channels in the middle and 
exhausted through the outlet near the outside temperature sensor. 
The pressure in the container was probed by two capacitance gauges 
with overlapping ranges, which were connected to the cryostat 
by about 10\,m of 0.5\,inch O.D. tubing.

Heat pulsing was performed by pulsing the power of the heater 
under a constant flow of helium. 
The conditions of the pulses were chosen to simulate the pressure rise 
during the proton pulse as observed at the PSI UCN source and
shown in Fig.~\ref{pressure-PSI}.
The maximum pulse duration at PSI is 8\,s, and 
the pressure increase is about 2\,Pa above the base line.
The baseline pressure observed at PSI  
is not considered to be the equilibrium vapor pressure of the sD$_2$ 
but is likely due to some small amount of $^4$He 
gas which leaked into the deuterium tanks.
The relative pressure rise observed during a pulse
is independent of the baseline value.
which was 2\,Pa in 2016,
and around 0.4\,Pa in 2017, as shown in Fig.~\ref{pressure-PSI}.
There is no temperature sensor inside the sD$_2$ at PSI, 
but thermal simulations show that the top of the
30\,dm$^3$ sD$_2$ bulk can reach temperatures up to 11\,K
by the end of the pulse, depending on pulse duration.

\begin{figure}[htb] 
\begin{center}
\resizebox{0.5\textwidth}{!}{\includegraphics{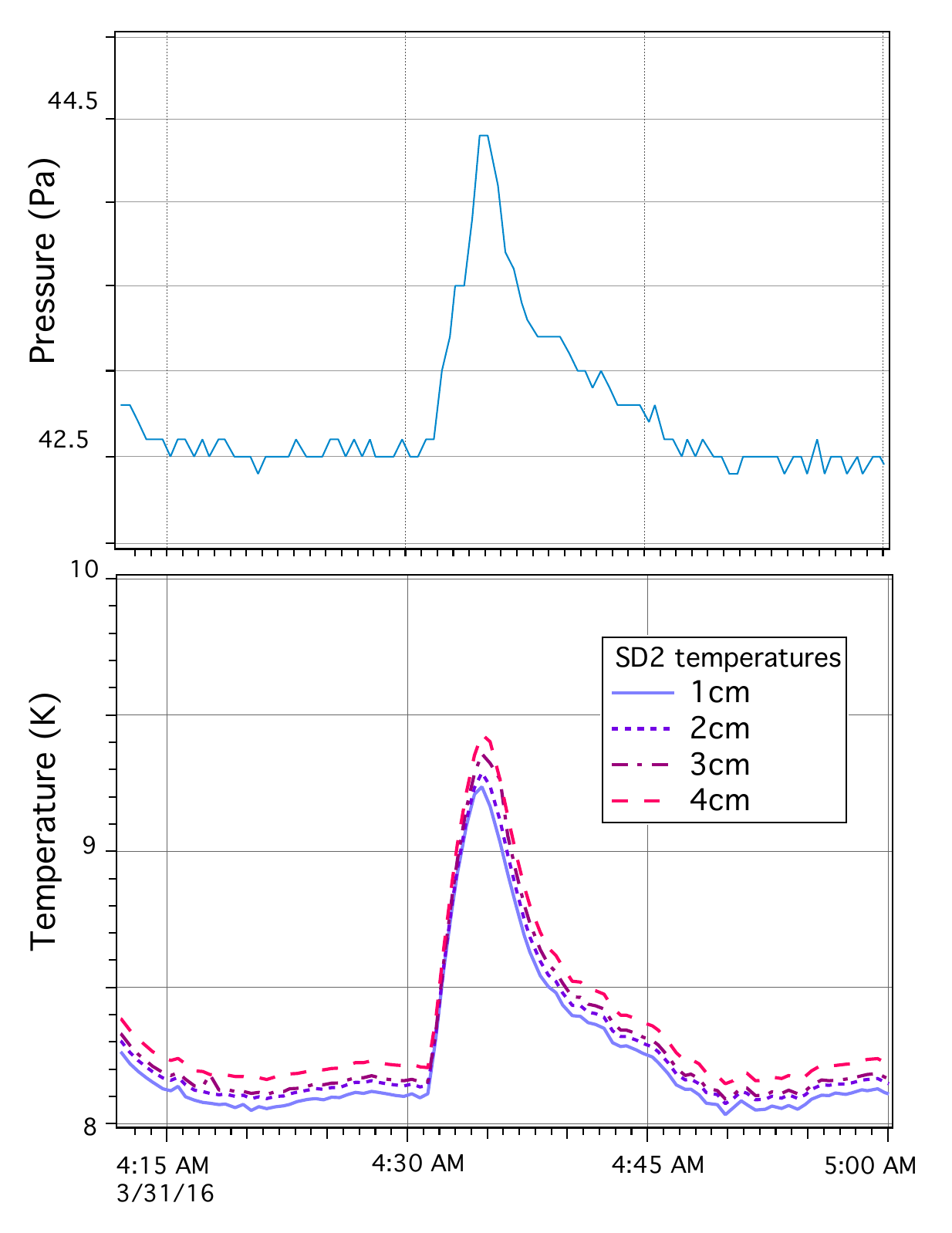}}
\caption{
Pressure (top) and temperature (bottom) rise 
in the deuterium container 
during single heat pulsing at NCSU.}
\label{pressure-NCSU}
\end{center}
\end{figure}

During the tests at NCSU with heat pulses the initial 
temperature of the solid was about 8.2\,K.  
The duration of the pulse was chosen to be 3\,min 
to provide a pressure rise of about 2 Pa.
%
During the heating, 
the temperature of the deuterium increased from 8.2\,K to 9.5\,K, 
and the container walls from 11\,K to 14\,K,  
as shown in Fig.~\ref{pressure-NCSU}. 
In total there were 54 heat pulses initiated.

\begin{figure}[htb] 
\begin{center}
\resizebox{0.5\textwidth}{!}{\includegraphics{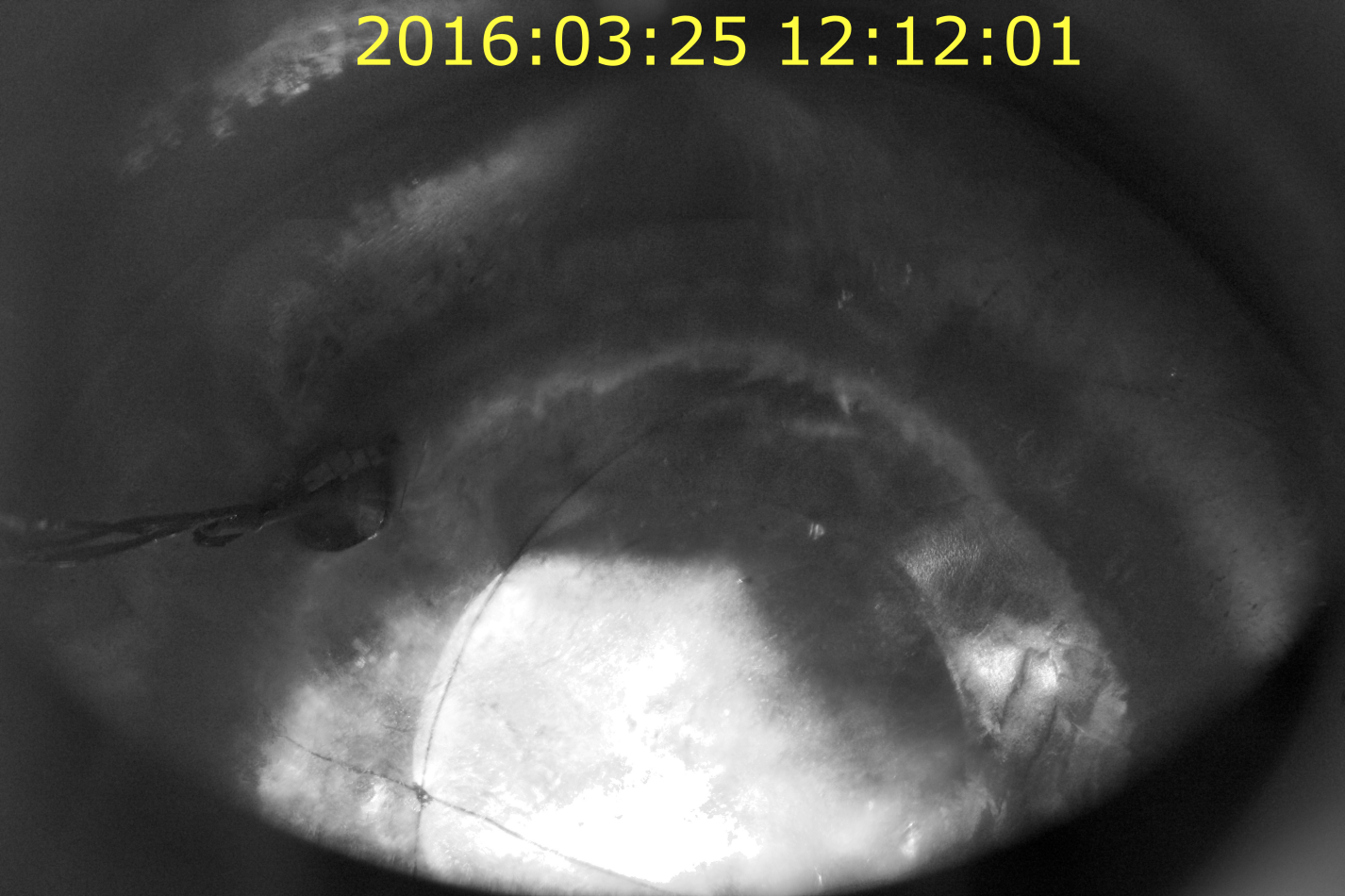}
}
\vspace*{0.3mm}
\resizebox{0.5\textwidth}{!}{\includegraphics{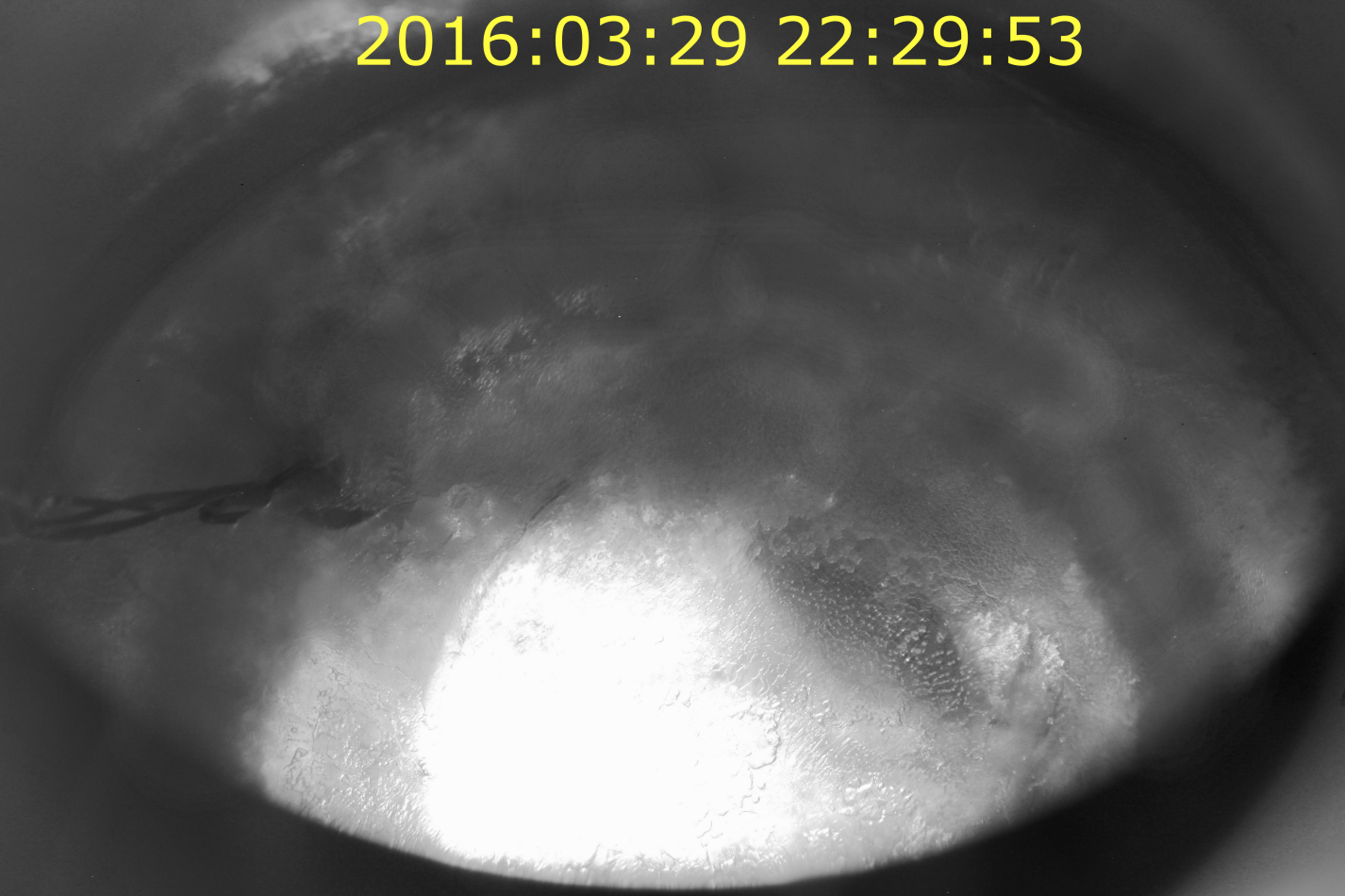}
}
\vspace*{0.3mm}
\resizebox{0.5\textwidth}{!}{\includegraphics{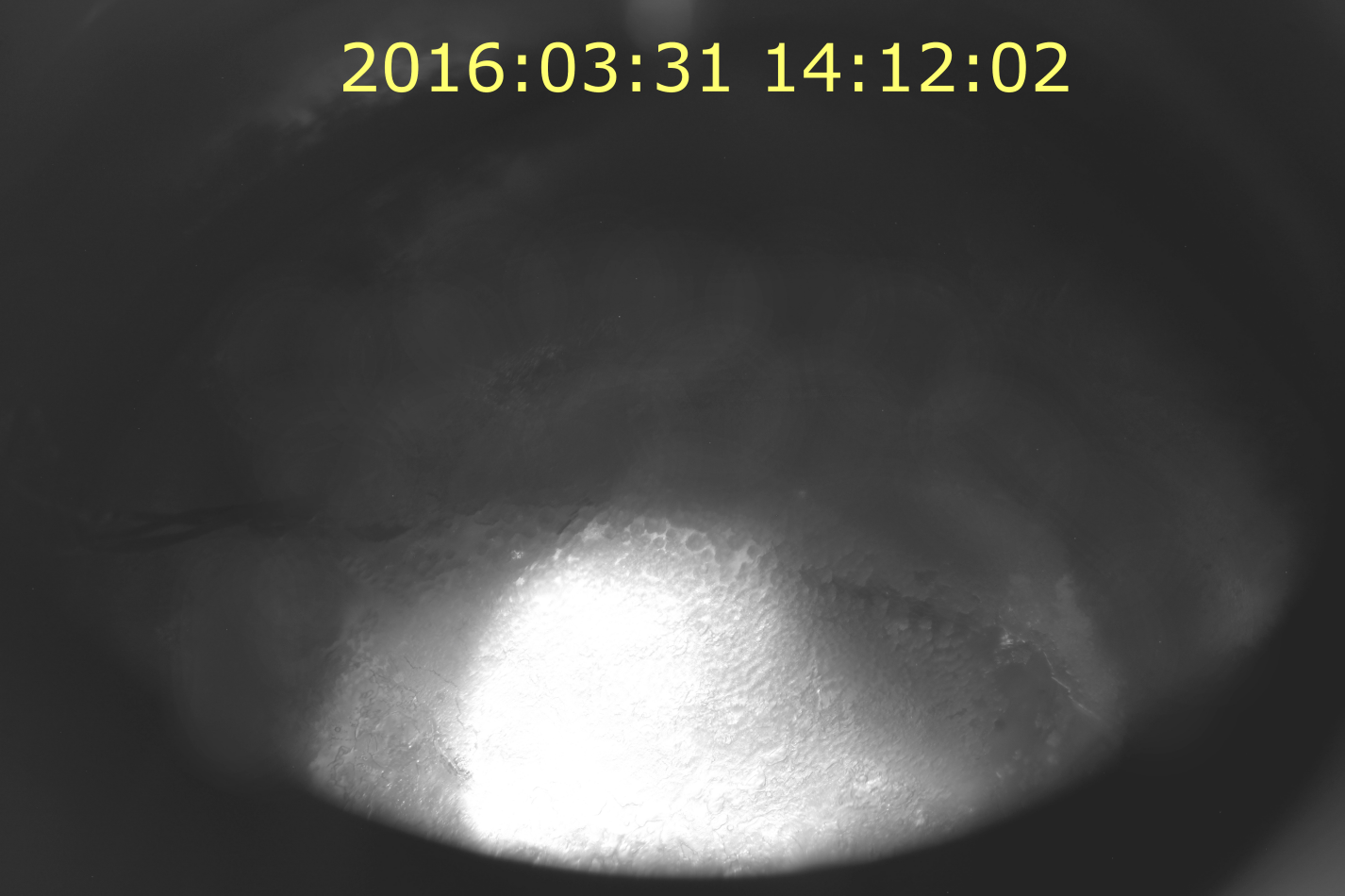}
}
\caption{Image of sD$_2$ 
top:
before heat pulsing after accidental 
He flow cycling;
center: during heat pulsing;
bottom: after heat pulsing.
}
\label{D2-crystal-heat}
\end{center}
\end{figure}

Two days before pulsing, the solid deuterium was grown from vapor.
Fig.~\ref{D2-crystal-heat}-top shows a clear crystal 
on the bottom of the cryostat and the temperature sensor located on the side.
The crystal was optically transparent, with just a tiny amount of frost 
after 4 accidental heat pulses due to an instability of the helium flow. 
During heat pulsing, the surface clearness deteriorated, 
while towards the edges one still can see the temperature-sensor holder 
through the frost layer.
(Fig.~\ref{D2-crystal-heat}-center).

\begin{figure}[htb] 
\begin{center}
\resizebox{0.5\textwidth}{!}{\includegraphics{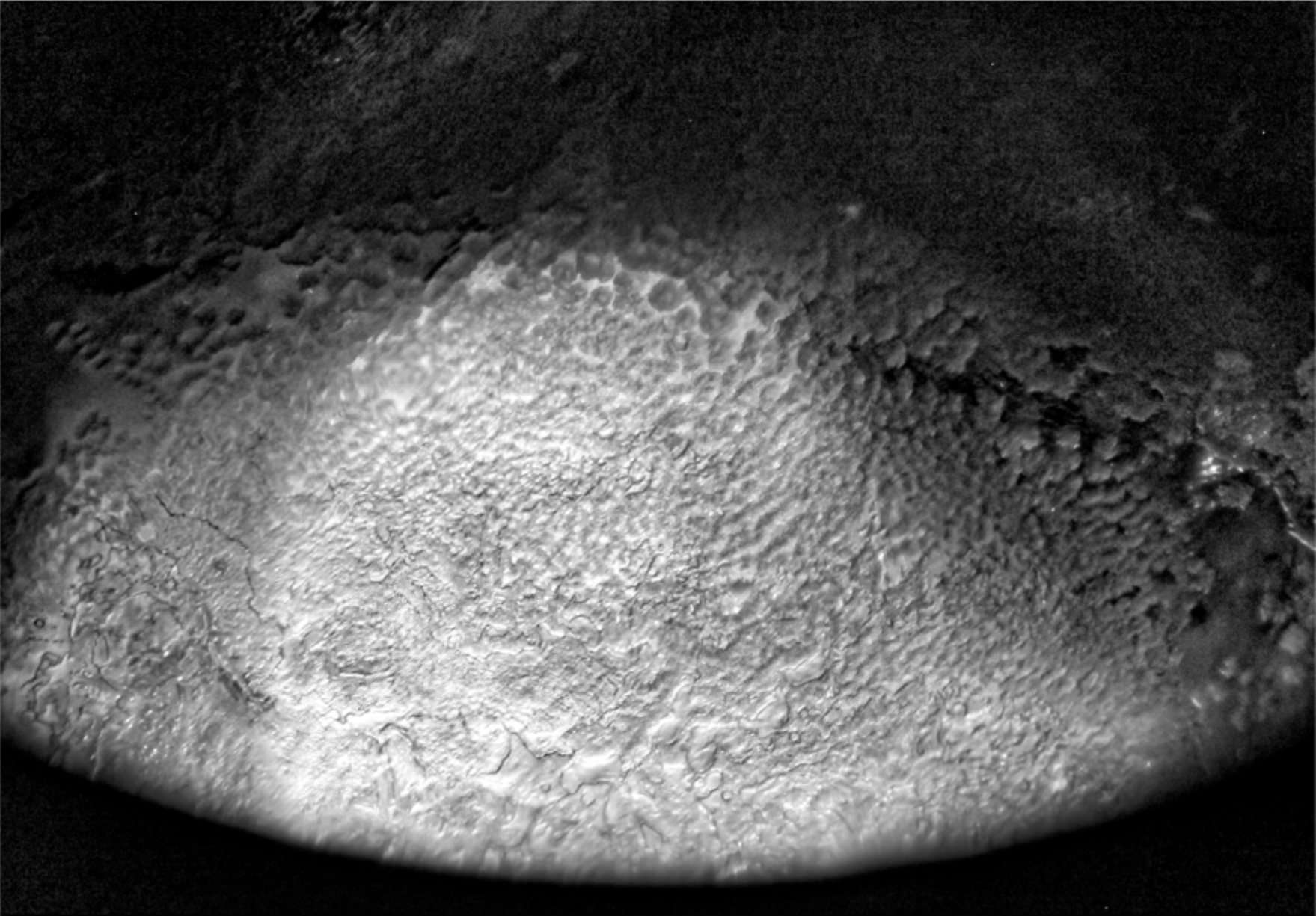}
}
\caption{Zoomed and post-processed image of the sD$_2$
surface after thermal pulsing.}
\label{D2-pic2}
\end{center}
\end{figure}

After heat pulsing, as shown in 
Fig.~\ref{D2-crystal-heat}-bottom,
the surface of the solid in the center looks very opaque 
and covered with some structured frost.  
Figure~\ref{D2-pic2} shows a zoomed and post-processed image of the frost. 
From the dimensions of the T-sensor holder (diameter 10\,mm)
we have estimated that the visible large structures are about 1\,mm in scale,
assuming that the vertical scales of structures  are similar to the horizontal ones. 



\section{Simulation and interpretation of experimental observations}
\label{simulation}

A combination of measurements during UCN source operation (Sec.~\ref{Sec:PSI}), 
optical observations (Sec.~\ref{optic-NCSU}),
and simulations, as described below, strongly support the hypothesis
of the surface origin of the UCN yield diminution.

In general, there are three ways that surface degradation can affect UCN transport. 
Firstly, a  top surface with a roughness on the scale of the UCN wavelength can change 
the shape of the neutron optical (Fermi) potential from a sharp step to a smooth profile. 
This can slightly affect reflection and transmission.


Secondly, the surface can broaden the angular distribution of UCN due to scattering from fluctuations in the density near the surface layer on scales much smaller than a typical crystal facet.  In this case, one typically describes surface interactions via two components: a  ``specular'' interaction with the average surface (facet) potential which obeys optical reflection laws (Eq.'s \ref{eq-reflectivity} and \ref{eq-transm}), and a component of ``non-specular scattering'' which results from density fluctuations about the flat facet surface (essentially independent of the elastic scattering mean free path in bulk sD$_2$).   
Non-specular scattering in UCN transport is often described by a Lambertian angular distribution with respect to the surface 
normal~\cite{Golub1991,Ignatovich1990,Zsigmond2018}, parametrized by a single constant, 
the average probability 
of non-specular scattering $\alpha$  
(see~\cite{Atchison2010}
and references therein for details on models for surface scattering for different distance scales and surface topologies). 

Finally, a layer of frost detached from the bulk will introduce an albedo effect resulting from scattering from randomly oriented crystal facets on the complex 
topology of a frost-covered surface~\cite{Kelly91}. 
When entering a facet, the neutron must have sufficient energy perpendicular to the facet to penetrate the potential barrier presented by the surface Fermi potential. 
When leaving the material, the UCN receives an energy boost in the direction perpendicular to the Fermi potential. Because the facets are oriented randomly, the result is increasingly like a diffusion problem as the number of encountered facets increases, with the flux through the frost more rapidly attenuated for low energy UCN. 

The last two effects were simulated by the NCSU and PSI groups 
using different UCN transport codes.
In the NCSU model, the effects of specular scattering on multiple layers
of frost covered with randomly oriented facets are explored, as well as the impact of a simplified,
Lambertian model of non-specular scattering at each facet face. The PSI model relies entirely on specular
scattering from facets.

The details of these models are rather different,
but the generic behavior is very similar: 
UCN interacting with successive layers of frost
experience an incremental loss in
the UCN flux observed above the frost layer,
with the largest losses for the lowest energy UCN.

In order to study the effect of repeated reflections at interfaces 
between sD$_2$ and vacuum, the simulations 
used the optical laws of reflection (R) and 
transmission (T) of UCN~\cite{Ignatovich1990}. 
The corresponding probabilities P$_T$ and P$_R$
are given by

\begin{equation}
\label{eq-reflectivity}
P_{R} = \left| \frac{\sqrt{E_{\bot}}-\sqrt{E_{\bot}-\Delta V_{F}}}{\sqrt{E_{\bot}}+\sqrt{E_{\bot}-\Delta V_{F}}}\right|^2,
\end{equation}

\begin{equation}
\label{eq-transm}
P_{T} = 1 - P_{R},
\end{equation}
where $\Delta V_{F}=\pm$102\,neV is the change in Fermi potential at the interface,  
$E_{\bot} = m v_{\bot}^2/2$, 
with $v_{\bot}$ being the velocity perpendicular to the interface,
the perpendicular part of the kinetic energy in vacuum 
or within the material before the interaction. 
%
These equations are valid for every energy $E_{\bot}$ below or 
above $\Delta V_{F}$. We neglected UCN losses since the UCN are only reflected a few 
times when passing the frost layer 
while the probability of loss per reflection is typically 
below $10^{-3}$~\cite{Bondar2017}.

\subsection{Simulations of rough and flat frost layers}



In simulations of
UCN transport through bulk sD$_2$ and frost layers the NCSU team explored
a number of effects and parameters  using a very simple source geometry, 
similar to the LANL source cryostat, 
but without long guide system external to the cryostat. 


All simulations were performed with a full energy spectrum:
an initial velocity distribution 
according to the low energy tail of the Maxwell-Boltzmann distribution was used:

\begin{equation}
\label{eq-maxwell}
P(v) dv = \text{const} \cdot v^2 dv,
\end{equation} 

%
with UCN velocities covering a range at the point of production in the deuterium
between 0 and 8\,m/s, more than covering the energy range of 
storable UCN in the guide system.  

UCN were produced homogeneously throughout the bulk sD$_2$ volume of the source,
with initial velocities directed isotropically in the top hemisphere,
and the yield, or probability of detection above the snow layers, was determined. 
For UCN transport through the bulk we assumed the sD$_2$ to be homogeneous, with
a total neutron life time (due to absorption and upscattering) of
$\tau_{sD_2}=$ 40\,ms and an elastic scattering mean free path 
= 4\,cm.
These values are adopted from LANL source characterization data, in which
Monte-Carlo is used to connect crystal parameters to the measured UCN yield
as a function of storage time in contact with the solid deuterium and the
amount of deuterium in the source (as described, 
for example, in~\cite{Morris2002,Saunders2004,LANL2013}).
These values should therefore be a reasonable starting point for our simulations,
but we also explore the impact of varying source conditions as a part of the studies presented in this section.

%
%

\begin{figure}[htb]
\begin{center} 
\resizebox{0.4\textwidth}{!}{\includegraphics{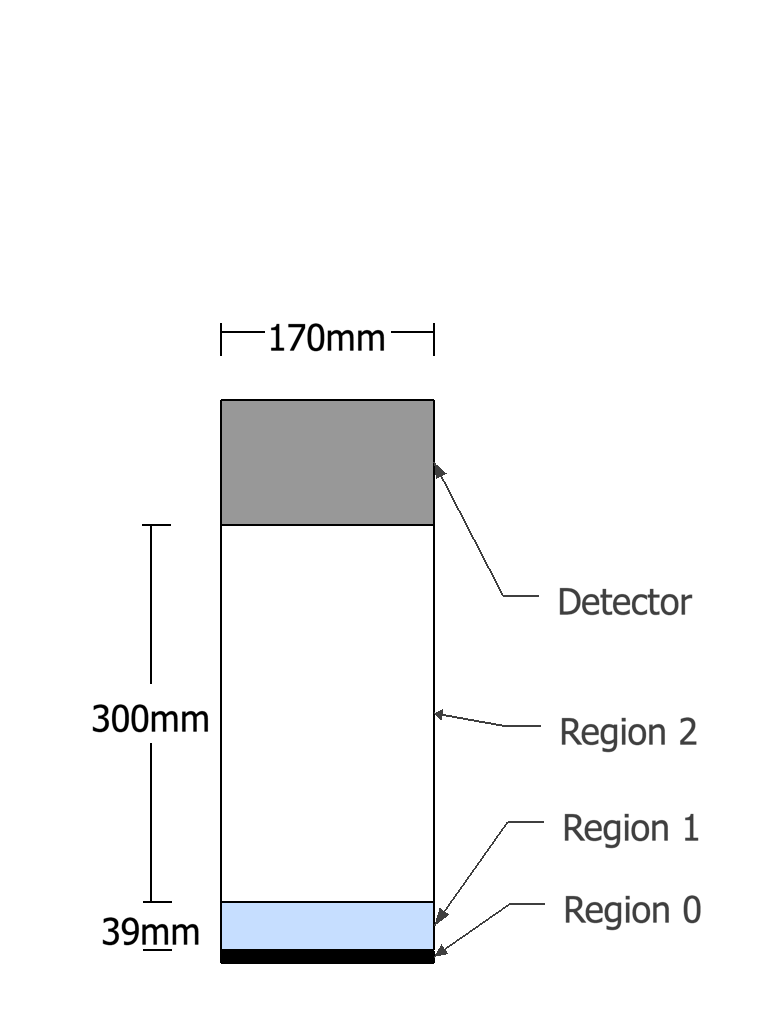}
}
\caption{
Straight geometry used for the simulations at NCSU, 
constructed with cylindrical components, corresponding dimensions of NCSU cryostat. 
Neutrons are produced in Region 1 and travel towards the detector. 
The bottom surface of the source (Region 0) was taken to be a $^{58}$Ni surface 
(with all neutrons entering the $^{58}$Ni assumed to be absorbed). 
The cylindrical walls of all regions were also assumed to be coated with $^{58}$Ni. 
}
\label{straightgeom}
\end{center}
\end{figure}

A simplified geometry, depicted in Fig.~\ref{straightgeom}, was used for the simulations
in this Section.  The sD$_2$ source was situated as a 3.9\,cm thick volume at the bottom
of a straight guide whose diameter matched the PULSTAR UCN cryostat guides (region 0).  The
bottom surface of the source was taken to be a $^{58}$Ni surface with Fermi potential $V_F = 335$\,neV
(with all neutrons 
which are not reflected from the $^{58}$Ni assumed to be absorbed).
The cylindrical walls of all regions were also assumed to be coated with $^{58}$Ni. 
A virtual
detector was placed above the frost layers, modeled as a 100\% absorbing surface.  
UCN incident
on this surface were counted as ``detected'' and their trajectories terminated.

The goal of the first simulation was to confirm that scattering
from multiple frost layers is needed to produce the experimentally observed reduction
in UCN yield. 
We compared the probability to detect UCN above the sD$_2$ bulk
for two cases assuming no frost layers above the surface and either
(i) a perfect flat surface for the solid, and (ii) 
a single faceted interface between the sD$_2$ and guide vacuum, with the facets
modeled by isotropically (described in more detail in the next paragraph) distributing
the direction of the surface normal at the point of interaction with the UCN. UCN reaching
a virtual detector above the frost layer, modeled as a completely absorbing disk, we
tallied as ``detected'' and contributed to the yield.  
The results
showed less than a 7\% difference in the neutron yield. Consequently, scattering from
a single layer of facets on the surface of the bulk  sD$_2$ can be ruled out as a cause
for decreasing UCN yield from the source.

\begin{figure}[htb] 
\begin{center}
\resizebox{0.5\textwidth}{!}{\includegraphics{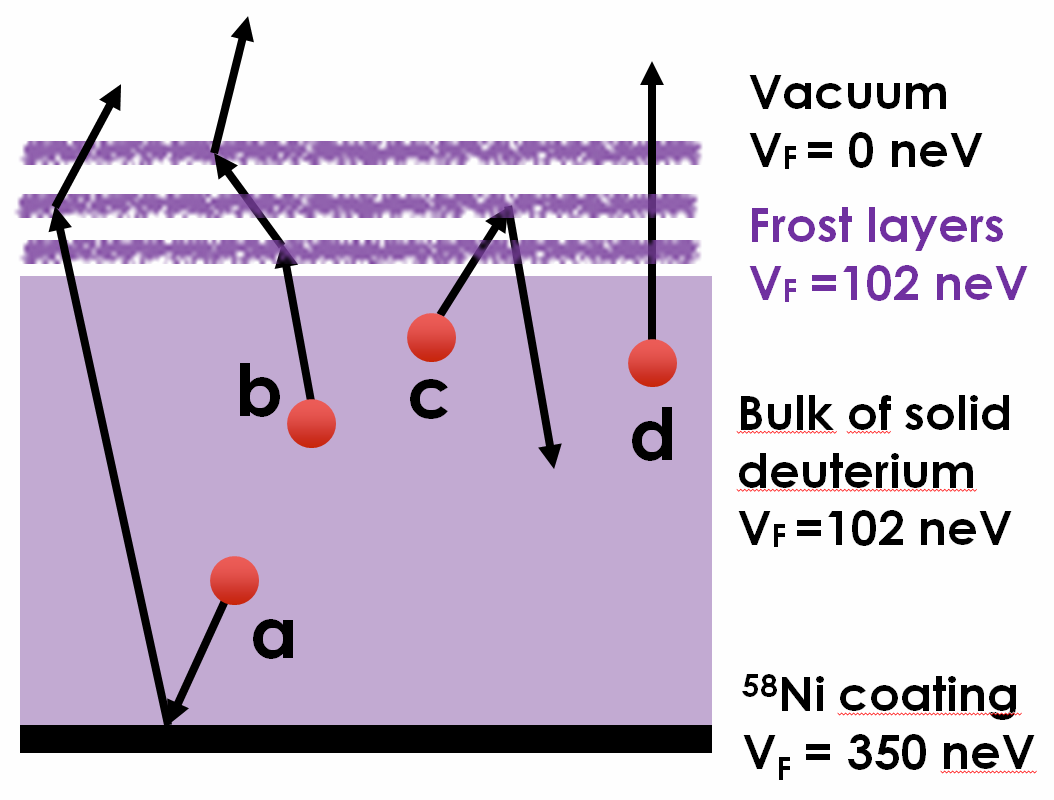}}
\caption{Frost layer model (NCSU): frost layers parallel and equal in area 
to the sD$_2$surface, separated by vacuum; 
UCN were isotropically produced in the bulk of 
the sD$_2$, with their velocities directed isotropically upward into $2\pi$ solid angle. 
Path examples: a -- neutron reflected at the bottom and single reflection from the frost; 
b -- double reflection from the frost;
c -- reflection back into the bulk and absorption;
d -- no reflection.}
\label{Frost-NCSU-model}
\end{center}
\end{figure}

As a next step, frost structures were added between region 1 and 2 
(see Fig.~\ref{straightgeom})
by alternating layers of vacuum and sD$_2$ ($V_{F}$=102\,neV) 
regions parallel to the solid surface, 
as shown in Fig.~\ref{Frost-NCSU-model}.  Scattering was assumed to
occur from both sides of each layer due to randomly oriented facets
and (when implemented) from non-specular scatter.  The surface normal was selected
isotropically, but the angular range of the normal was constrained such that reflection
at a boundary layer prevented the UCN from proceeding vertically into the next frost (or
vacuum) layer.  While simplistic and not exactly reproducing the effect of random orientation 
of ``facets'' of sD$_2$ frost crystals,  
this model allows studies of the dependence of successive UCN scattering 
from vacuum/frost and frost/vacuum interfaces on conditions 
in the bulk sD$_2$ and the frost layers.

The thickness of each sD$_2$ and vacuum layer was chosen arbitrarily to 
be 10 microns to be roughly consistent with horizontal dimensions of optically observed 
features when many layers are present (features down to fractions of a millimeter
were evident after heat pulsing).  
Losses and elastic scattering
within the layers were confirmed to be negligible for the
full range of UCN lifetimes and elastic scattering mean free paths 
considered for the bulk deuterium (presented below).   

\begin{figure}[htb]
\begin{center}
\resizebox{0.5\textwidth}{!}{\includegraphics{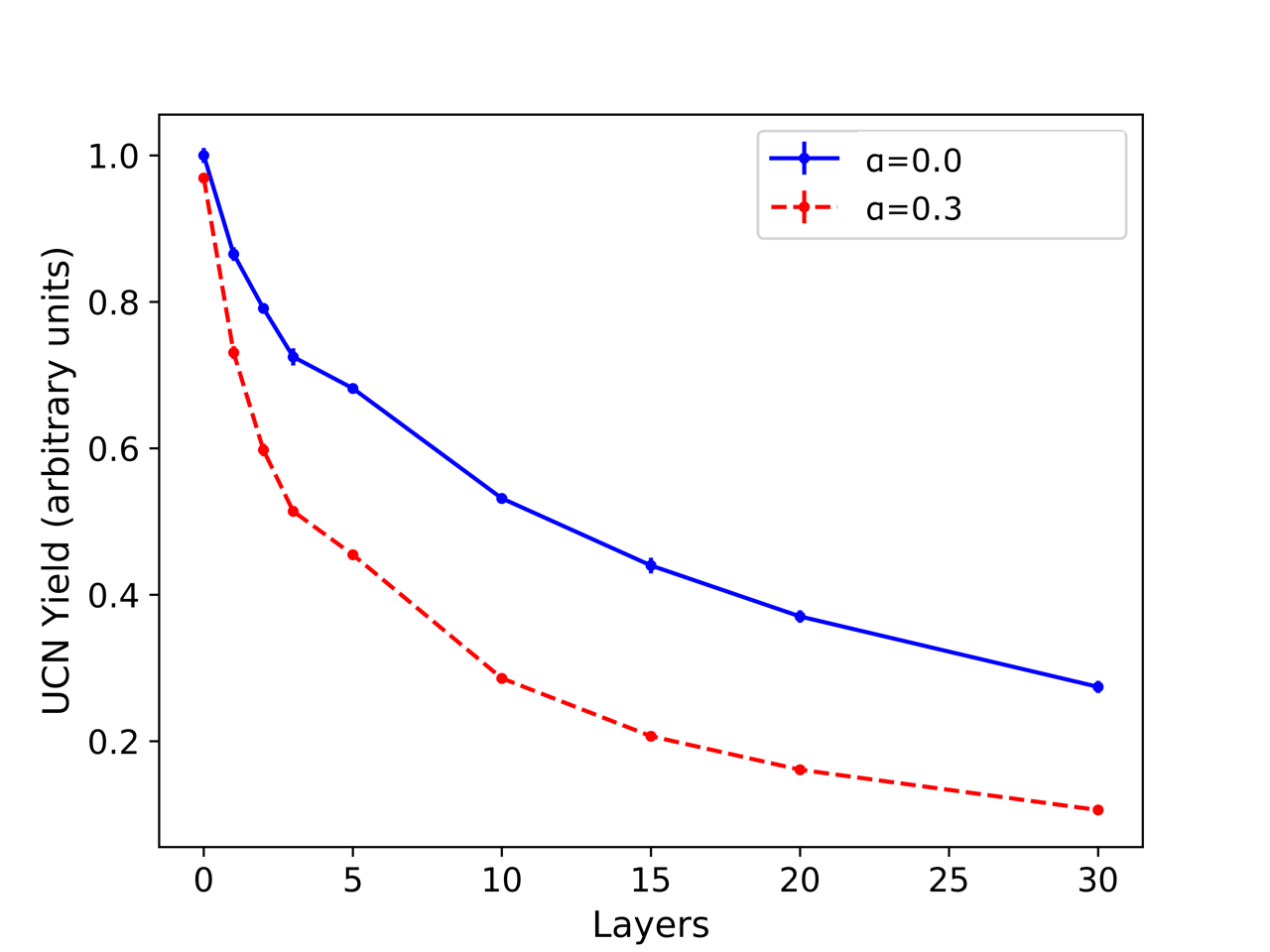}} 
\caption{UCN yield dependence on the number of frost layers for two different non-specular scattering probabilities: 
$\alpha = 0$ (perfectly specular) and $\alpha = 0.3$ (30\% probability of non-specular scattering).}
\label{correlations}
\end{center}
\end{figure}

In the first study with frost, the correlations of the UCN yield with the number 
of sD$_2$ layers was calculated by alternating 10 micron vacuum and deuterium layers. 
The results shown in Fig.~\ref{correlations} demonstrate an exponential attenuation
of the neutron yield, with pure specular scattering from facet surfaces (this corresponds
to $\alpha =0$ for our crude model of non-specular scattering).  We also attempt to show the
potential impact of non-specular scattering by adding the effects of rough surfaces to our
randomly oriented facet surfaces' with a second curve with $\alpha = 0.3$, corresponding
to a rough surface.  
As discussed in Sec.~\ref{simulation}, 
non-specular scattering results from roughness on the
surface of a given facet.  Of course, the surface profile for frost facets is an unknown function
of the vapor flux, temperature variations, and crystal history, for example, but to clearly
illustrate the impact of this scattering, we selected a value of $\alpha = 0.3$. For reference,
for highly polished guide materials, $\alpha$ is of order 0.03 or smaller~\cite{Atchison2010}.
Non-specular scattering
was described by a Lambertian angular distribution with respect to the surface normal. The angular
distribution after non-specular scattering was also taken to be Lambertian.  Because our model
of non-specular scattering does not depend on the energy of the UCN incident on the surface (but
specular reflection from randomly oriented facets more strongly affects the lowest energy UCN)
and our facet model features a restricted range of facet orientations, it is evident that non-specular
scatter at frost surfaces can increase the attenuation of the UCN flux, but
does not change the qualitative picture of successive layers of frost incrementally reducing the
UCN yield.  Given that purely specular scattering appears to capture the essential physics of transport
through the frost layers, we restrict our analysis from this point forward to specular scattering ($\alpha = 0$).

Next we studied the dependence of the yield on the frost layer Fermi potential (which is proportional to the frost density).  Fermi potentials of 100\%, 66\% and 33\% of that for bulk sD$_2$ were simulated.  We found, as expected, that the yield losses scaled with the Fermi potential.

\begin{figure}[htb]
\begin{center}
\resizebox{0.5\textwidth}{!}{\includegraphics{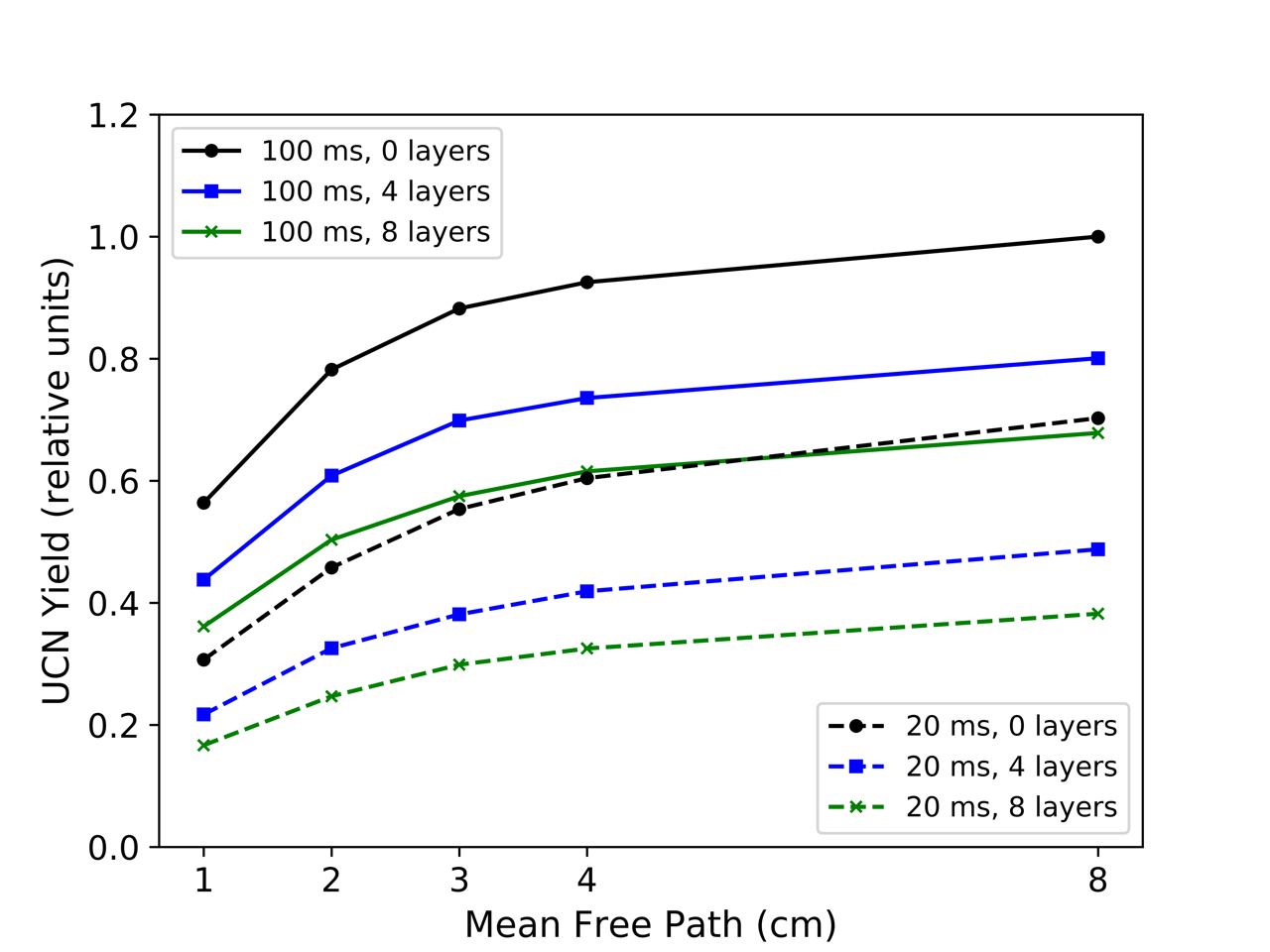}} 
\caption{UCN
  yield dependence
  on the number of frost layers 
versus UCN mean free path for two different UCN life times 
in the bulk of sD$_2$ crystal, which is 5\,cm thick 
(largest value normalized to unity). Only specular reflections are considered.}
\label{BulkCorrelations}
\end{center}
\end{figure}

Since the main effect of frost is reflecting neutrons back to the bulk, 
the UCN yield should be dependent (at some level) on the bulk properties of sD$_2$ 
We have explored this effect by varying neutron life time and mean free path
in the bulk sD$_2$.
Results for two values of $\tau_{sD_2}=$ 20\,ms and 100\,ms 
are presented in Fig.~\ref{BulkCorrelations}. 
For $\tau_{sD_2}=$ 100\,ms,
the UCN yield
through 4 layers of frost 
is higher then in the case of no frost 
with $\tau_{sD_2}=$ 20\,ms. Note that, because of the very small volume
in the frost layers, no dependence was observed on bulk sD$_2$ properties
in the frost layers themselves, only in the sD$_2$ source volume.

Changes in the bulk sD$_2$ properties may explain the observed scatter in the starting source performance 
after each melt-and-refreeze, and has some influence on the efficiency 
with which successive layers reduce the source output and and produce spectral hardening.
%
The relative impact on source performance of the bulk sD$_2$ properties, together with the 
integrated beam power, volume of solid deuterium in the source, 
physical configuration of the bulk sD$_2$ (how much deuterium ends up on the side walls of the 
source below the flapper valve, for example) after a melt and refreeze, 
the vapor density above the sD$_2$ and the cryogenic operating conditions 
for the source are under continuing investigation.
These variables can also explain the spread in the attenuation slopes in the LANL data,
but they also make exact simulations of the LANL source more challenging,
by adding additional free parameters.
Note that, although the UCN lifetime in the bulk varied rather slowly 
over the course of many months of LANL running,
%
%
%
each melt and refreeze can, in principle, produce large variations 
in the effective elastic scattering mean free path in the bulk  
as reported in various
studies~\cite{Serebrov2001b,Atchison2005b,Atchison2011}.

\begin{figure}[htb]
\begin{center}
\resizebox{0.5\textwidth}{!}{\includegraphics{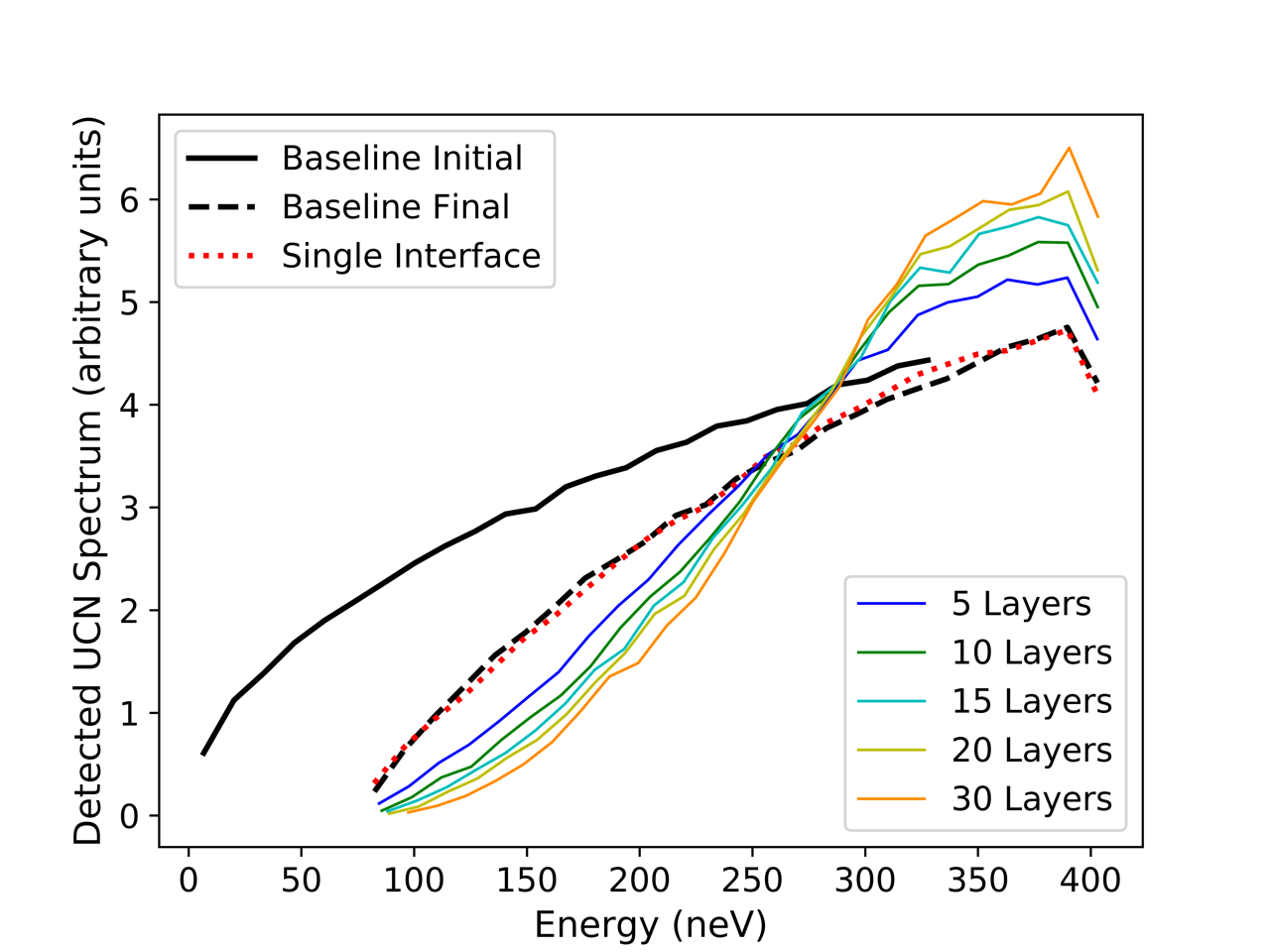}} 
\caption{The UCN energy spectrum shape vs the number of 
sD$_2$ layers. 
``Baseline initial'' reproduces the initial spectrum 
from eq.(\ref{eq-maxwell}), 
``baseline final'' indicates the distribution in the detector without frost; 
this spectrum starts at the boost energy of 
the sD$_2$ diminished by the gravitational potential 
at the detector. Only specular surface reflections on the frost are considered }
\label{NCSU-simul-Energy}
\end{center}
\end{figure}

Another simulation was performed to study the effect of the
frost on the detected
neutron energy spectrum 
while varying the number of layers.
Results are shown in Fig.~\ref{NCSU-simul-Energy}. 
The spectra are normalized to unit area for better comparison
of their shape.  

Two conclusions can be drawn. 
Firstly and most importantly, there is clear evidence of spectral hardening with increasing number of frost layers.
Secondly, spectral hardening effects, which were experimentally observed at PSI and LANL,
cannot be explained by a single rough interface between deuterium and vacuum.  
It should be noted that, in order to track the effect of successive interactions with frost layers, the frost layers cover the entire guide cross section in this first model and hence represent a limiting case of sorts. 
This simple model for interactions with the bottom surface of a frost layer 
(where UCN either enter the frost layer or are reflected down towards the sD$_2$ source, 
but reflections from the ``edge'' of a frost feature which scatter UCN upward are not possible) results in gradual losses in
UCN yield
and correlated ``hardening'' of the spectrum, as observed for both LANL and PSI sources. 
Of course, voids or ``holes'' in the frost layers or reduced layer density can be incorporated into this model, accommodating a continuous variation of the impact of frost layers from no frost at all to many completely covering layers, 
but the characteristic, correlated spectral hardening and
reduction in yield should always be present.
 
A more flexible model of the PSI source presented in the next section incorporates isotropic, random reflections from each frost surface, and variable coverage of scatterers permits tuning the impact of the frost layers on transmission and spectral hardening to accurately reproduce the measured performance of the PSI source.


\subsection{Simulation and interpretation of the PSI measurements}
\label{simulation-psi}

In order to examine the fraction of UCN transmitted 
through a frost layer, the reflection from frost (snow) particles was 
modeled in 3D by using the MCUCN code~\cite{Zsigmond2018}. 
The model describes the effect of reflections due to the optical potential step between
sD$_2$ and vacuum as described by the quantum reflection and 
transmission laws above, eqs.~(\ref{eq-reflectivity})--(\ref{eq-transm}).

\begin{figure}[htb] 
\begin{center}
\resizebox{0.45\textwidth}{!}{\includegraphics{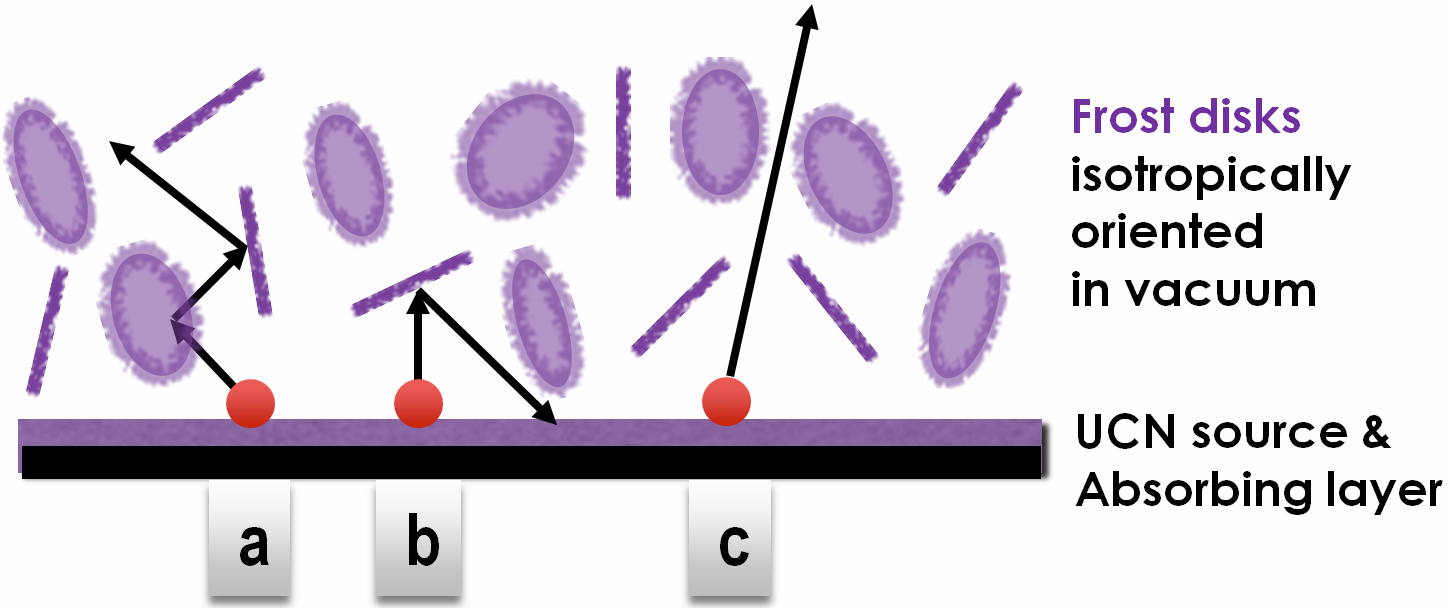}}
\caption{Frost ''disks'' model (MCUCN/PSI): thousands of frost disks which were randomly 
oriented and small in area compared to the sD$_2$ surface.
Here only a small number of discs is shown for illustration purposes. 
UCN were produced at the source surface and backed with a 100\% absorbing layer. 
Path examples: 
a -- double reflection from two disks; 
b -- back-reflection; 
c -- free escape.}
\label{Frost-PSI-model}
\end{center}
\end{figure}

The geometry consisted of a large number of small sD$_2$ disks distributed 
isotropically in orientation
and homogeneously over the source surface, and several disk-radius units 
above the bulk, see Fig.~\ref{Frost-PSI-model}.
The vertical density distribution of the disks turned out to be irrelevant. 
The main assumption to the disk thickness was that 
(i) it is thin enough to neglect 
nuclear absorption or up-scattering since the mean-free-path associated 
with these losses is much larger; 
(ii) it is thicker than several UCN wavelengths, so that 
assuming sD$_2$ bulk density,
an optical potential 102\,neV can be used for the material without allowing for quantum tunneling. 

Multiple reflections within the disks between two faces were neglected, 
since this is a second order effect. 
Only specular reflections were considered because the isotropic angular 
distribution of the disks already accounts for a perfect 
diffuse redistribution after reflection. 
The number of disks was varied up to several thousand. 
The radius of the disks was varied in order to show that only the total surface of all disks normalized to the cross section of the sD$_2$ converter is relevant. 
Doubling the disk radius was found equivalent to a four times lower 
number of sD$_2$ disks. 

In order to obtain the net transmission of only the disks layer, 
the virtual UCN source (a diffuse emitter) was set to perfectly 
absorb back-reflected particles. 
The effect of the energy boost from a perfectly flat sD$_2$ surface
was also compared to the case of an isotropically boosting emitter with a rough surface. 
The latter is equivalent to the case of no vertical boost but a final energy higher by 102\,neV. 
The lateral boundary of this geometry was a cylindrical guide. 
This was set to have an ideally high optical potential, 
and the loss and diffuse reflection parameters were set to zero.

\begin{figure}[htb] 
\begin{center}
\resizebox{0.5\textwidth}{!}{\includegraphics{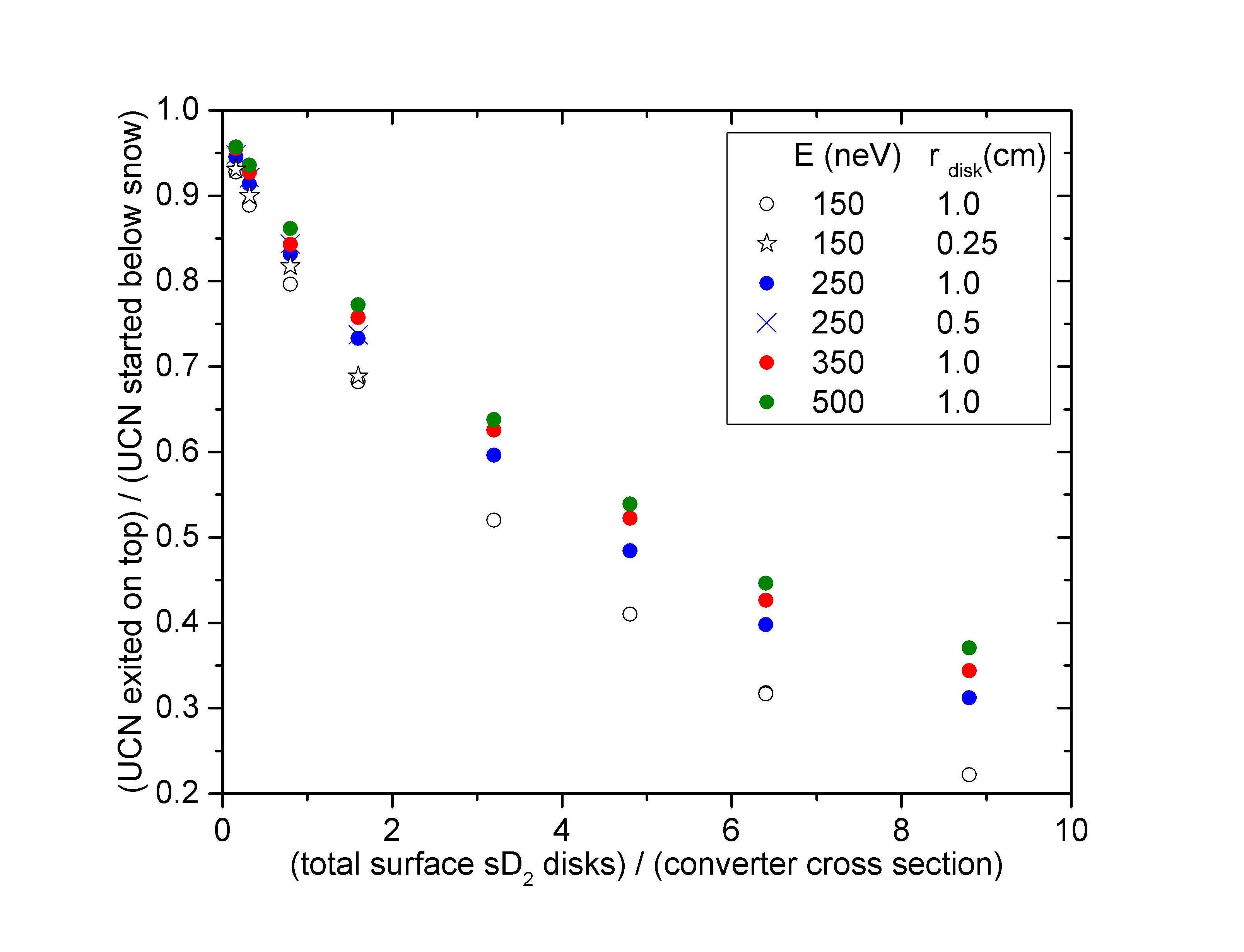}
}
\caption{UCN transmission of a sD$_2$ layer formed by isotropically oriented 
disks having 102\,neV optical potential as simulated with the MCUCN code.}
\label{MCUCN-sD2-disks-transmission}
\end{center}
\end{figure}

The calculated transmission is shown in Fig.~\ref{MCUCN-sD2-disks-transmission} 
for various neutron energies as a function of the total sD$_2$ frost surface.
The UCN transmission of the sD$_2$ disks layer is given by the ratio of the number of transmitted to incident UCN. 
%
The horizontal axis is the ratio of the total surface of all disks, and the cross section of the virtual source (and the cylinder guide). 
These results show that the UCN transmission 
decreases with increasing frost layer surface.
The energy dependence indicates that such a reflecting layer
attenuates slower UCN more strongly.
%
Even a low surface ratio 
can cause a large UCN loss 
only due to back-reflections from randomly oriented interfaces 
between sD$_2$ and vacuum. 
%

%

These simulation results reproduce the magnitude of the decrease 
in UCN output observed after several days at the PSI UCN source.  
The decrease to 0.7 
would be equivalent to a simulated surface ratio 
of about 2 (see Fig.~\ref{MCUCN-sD2-disks-transmission}). 
Furthermore, this corresponds to the amount of deuterium
estimated in the conditioning treatment,
equivalent to a layer thickness of approximately 500-1000\,nm. 
This allows the disk layer thickness of sD$_2$ 
to be 250 to 500\,nm,
i.e. about 4 wavelengths for 150\,neV UCN. 
This thickness is large enough for the formation of a totally reflecting optical potential. 
The consistency of the measured and simulated values indicates 
that a frost scenario is realistic.

This model can explain the decrease 
shown in 
Fig.~\ref{yield-decline-LANL-PSI},
Fig.~\ref{yield-decline-PSI-weekend}
and in Fig.~\ref{yield-ratio-PSI}. 
UCN extracted in West-2 are emitted by the sD$_2$ with energies much 
larger than its optical potential and are consequently 
less affected by the frost layer.

\section{Summary}

The existing
pulsed UCN spallation sources based on sD$_2$ converters
suffer from a significant, energy dependent decrease in UCN intensity 
with increasing number of pulses
after initial preparation of the sD$_2$.
This preparation can be melting and refreezing of the total material
(at LANL, with change of the sD$_2$ bulk properties) 
or a special heat treatment of the solid called ``conditioning'' 
(at PSI, likely without change of the sD$_2$ bulk properties).

We have presented energy dependent UCN intensity measurements 
at the pulsed PSI and LANL UCN sources, 
which show the lower the neutron energy, the stronger the attenuation.
Together with correlated measurements of temperatures and D$_2$ vapor pressures
during pulse operation and during conditioning (at PSI),
all of the data pointed towards the build-up of frost 
layers on the sD$_2$ surface as most realistic explanation for reduced intensity with larger pulsed heating. 
Pulses cause sublimation of D$_2$ with resublimation at a different temperature 
during and after the pulse.

Optical observations of sD$_2$ at the
PULSTAR UCN source setup have been performed by applying heat pulses with a resistance heater.  
The heat pulses were adjusted to have profiles similar to those measured from the beam 
heating at the operating UCN sources. 
These observations clearly demonstrate the adverse effects 
on the surface quality of the sD$_2$ from the pulsed heating, which turns from
completely optically transparent to opaque and structured.

Transport simulations through the frost layers were performed 
by both, the NCSU 
(for a simple geometry similar to LANL source cryostat only) 
and PSI teams 
(using a more realistic frost model).
The NCSU simulations were performed to explore general parameter 
dependencies of
the UCN yield when frost is present,
including dependence of frost attenuation on the crystal bulk properties. 
The PSI simulation was able not only to explain the observed count rate decrease
but also to extract realistic frost parameters.  

The UCN measurements at PSI are very well described by the PSI model. 
The frost layers are simulated by isotropically 
oriented sD$_2$ disks accumulating on the D$_2$ surface. 
UCN can undergo reflections from all surfaces on their path. 
The simulation shows that
the UCN transmission through the layers of disks depends on the
UCN kinetic energy and the surface ratio of all disks
relative to the converter area and that the transmission is independent of size
and shape of the disks. 
As the total surface area of the disks grows 
(corresponding to more pulses and longer operating time) 
the UCN transmission is decreasing due to increased
albedo reflection.

In conclusion, the complete set of UCN counting data, 
optical observations and different simulation models 
support surface frost formation as the explanation 
for UCN yield decline with time at pulsed UCN sources. 
This deterioration is correlated to the pulsed heat deposition, which leads to material transport and reorganization 
on the sD$_2$ surface.
A better understanding of the cause may allow further improving the 
conditioning process or even
controlling the sD$_2$ conditions under pulsed operation 
in a way that avoids the build-up of frost layers.

\section*{Acknowledgments}
This work was supported in part by the US Department of Energy 
under Grant No.\ DE-FG02-97ER41042 and 
the US National Science Foundation under grant number NSF-1615153.
PSI acknowledges the support by the Swiss National 
Science Foundation Projects 
200020\_137664, 
200020\_149813, 
and 
200020\_163413, 
support by the proton accelerator operations and UCN source operations groups,
support by P.~Erismann, M.~Meier and C.~Zoller, 
and access to the computing grid infrastructure PL-Grid~\cite{PLGrid}.
We acknowledge Jan Bacca for helping proof read the paper.

\bibliographystyle{unsrt}

\end{document}